\documentclass[12pt]{elsarticle}

\usepackage{amsmath}
\usepackage{amssymb}
\usepackage{graphicx}
\usepackage{hyperref}
\usepackage{color}
\usepackage{multicol}
\usepackage{calrsfs}
\usepackage{algorithm}
\usepackage{algorithmic}
\usepackage{cancel}
\usepackage{multirow}
\usepackage{caption}
\usepackage{subcaption}
\usepackage{upgreek}
\usepackage[algo2e]{algorithm2e} 
\usepackage{hyperref}
\begin{document}
	
	\vspace{.2in}\parindent=0mm
	
	\begin{flushleft}
		{\bf\Large { Continuous Boostlet Transform  and   \vspace{.1in}  Associated Uncertainty Principles }}
		
		\parindent=0mm \vspace{.3in}
		{\bf{  Owais Ahmad and Jasifa Fayaz}}
	\end{flushleft}
	{{\it Department of  Mathematics,  National Institute of Technology, Hazratbal, Srinagar -190006, Jammu and Kashmir, India. E-mail: $\text{siawoahmad@gmail.com;jasifaitoo058@gmail.com}$}}
	
	\parindent=0mm \vspace{.1in}
	{\bf{Abstract:}}~~The Continuous Boostlet Transform (CBT) is introduced as a powerful tool for analyzing spatiotemporal signals, particularly acoustic wavefields. Overcoming the limitations of classical wavelets, the CBT leverages the Poincaré group and isotropic dilations to capture sparse features of natural acoustic fields. This paper presents the mathematical framework of the CBT, including its definition, fundamental properties, and associated uncertainty principles, such as Heisenberg's, logarithmic, Pitt's, and Nazarov's inequalities. These results illuminate the trade-offs between time and frequency localization in the boostlet domain. An  example of an exponential function highlights the CBT's adaptability.  With applications in radar, communications, audio processing,and seismic analysis, the CBT offers efficient time-frequency resolution, making it ideal
	for non-stationary and transient signals, and a valuable tool for modern signal processing.

	\parindent=0mm \vspace{.1in}
	
	\parindent=0mm \vspace{.1in}
	
	\parindent=0mm \vspace{0.2in}
	{\bf{1. Introduction}}

	Signal processing, a foundational discipline within applied mathematics, continues to shape modern science and engineering through its powerful ability to represent, analyze, and manipulate signals across continuous and discrete domains \cite{aa}. Among its most dynamic subfields, audio signal processing has profoundly transformed how humanity captures, perceives, and reproduces sound. It underpins technologies that not only enhance auditory experiences but also bridge computational mathematics, acoustics, and human perception. By enabling precise spectral control, noise suppression, and perceptual optimization, it serves as both an analytical framework and a creative tool for realizing complex auditory environments. The field witnessed a rapid evolution in the 1960s with the emergence of digital computers capable of translating Fourier’s theoretical insights into practical computation \cite{mbc}. The development of the Fast Fourier Transform (FFT) marked a paradigm shift, drastically reducing computational complexity and paving the way for real-time audio and acoustic analysis. Subsequent innovations, including the image source method for simulating realistic room reflections \cite{jd} and the revolutionary introduction of acoustic holography in the 1980s \cite{ej}, expanded the analytical reach of acoustics by enabling contactless investigation of vibroacoustic phenomena. Building on these milestones, Berkhout and his colleagues in the late 1990s ingeniously linked principles from seismic exploration to the acoustic modeling of enclosed spaces \cite{adj, adjb}. Presently, computational acoustics thrives on numerical strategies—such as finite element analysis, finite-difference-time domain formulations, spectral element methods, and discontinuous Galerkin schemes—that convert complex differential systems into computationally tractable algebraic representations.

\parindent=8mm \vspace{0.1in}
In parallel, the rise of multi-scale signal representations has redefined the analytical boundaries of signal processing. Wavelet theory, in particular, has emerged as a transformative tool for uncovering localized features across scales and dimensions \cite{jfa}. Yet, as the complexity and dimensionality of data have grown, traditional wavelets—limited by their isotropic structure—have faced challenges in effectively describing directionally dependent singularities \cite{jm}. This realization led to a new generation of geometric multiscale transforms designed to handle highly anisotropic and structured data. Constructs such as ridgelets \cite{edr}, curvelets \cite{edc, dmc}, contourlets \cite{mnd}, bendlets \cite{gs}, shearlets \cite{dwg}, grouplets \cite{sm}, and wavelet packets \cite{ngk} have extended the expressive power of signal representations. In acoustics, curvelets \cite{el} and wave atoms \cite{ll} have proven particularly adept at describing the sparse structure of wave propagators in free space—faithfully capturing the natural evolution of wavefronts as they propagate along characteristic trajectories governed by Hamiltonian flows. Demanet and Ying \cite{ll} demonstrated that these transforms achieve optimal sparsity for Fourier integral operators and Green’s functions of the wave equation, with their anisotropic essential support scaling as \(\sim 2^{-aj} \times 2^{-j/2}\),  with $a$  ranging between 1/2 and 1. Nevertheless, fully extending such directional and multiscale representations to the domain of full space-time analysis remains an open and compelling challenge \cite{tgtb,fm, cey, em}.

\parindent=8mm \vspace{0.1in}
While the above multiscale tools were initially tailored for image and geometric data, they were not fundamentally designed to capture the intrinsic wave dynamics that unfold across both space and time. This limitation was recently addressed by Zea et al. \cite{emj}, who introduced the Boostlet Transform (BT) in 2024—a transformative framework for analyzing acoustic waves directly within the space-time continuum. By integrating the symmetries of the Poincaré group with isotropic dilations, the Boostlet Transform elegantly encodes sparse, physically meaningful structures within acoustic fields. It bridges geometry, physics, and computation in a unified representation—one that naturally aligns with the causal and relativistic principles governing wave behavior.

\parindent=8mm \vspace{0.1in}
The Boostlet Transform remains in its formative stage, yet it carries remarkable potential to redefine how wave phenomena are analyzed and interpreted. This work seeks to fortify its theoretical foundations while inspiring broader exploration into its analytical capabilities. Specifically, we establish the fundamental mathematical properties of the Continuous Boostlet Transform (CBT) and derive a new class of uncertainty inequalities that govern its structure—relationships that, to the best of our knowledge, remain unexplored in existing literature. These results not only reveal intrinsic limits on the transform’s precision and localization potential but also provide a deeper understanding of the interplay between spatial resolution and frequency concentration. Beyond theory, this framework opens the door to novel applications that transcend conventional signal analysis, promising impact across acoustics, wave physics, and emerging data-driven paradigms where space-time modeling is essential.

\parindent=8mm \vspace{0.1in}
The paper is organized as follows. Section 2 presents the mathematical formulation of the Continuous Boostlet Transform (CBT) along with its essential properties. Section 3 establishes the associated uncertainty principles and explores their theoretical implications. Section 4 offers practical examples that illustrate the transform’s unique capability in analyzing diverse classes of signals. Finally, Section 5 discusses potential applications and outlines promising directions for future research inspired by the transformative potential of the Boostlet framework.

	\parindent=0mm \vspace{0.1in}
	{\bf {2. Continuous Boostlet Transform and its Fundamental Properties}}
	
	\parindent=0mm \vspace{0.1in}
	In this section, we establish some important results related to boostlet transform and  various fundamental properties of the continuous boostlet transform viz., linearity, quadratic homogeneity , translation, scaling and reflection. We first recall the definition of continuous boostlet transform in the function space $L^2(\mathbb {R}^2) $.
	
	\parindent=0mm \vspace{0.1in}
	\textbf{Definition 2.1.} \cite{emj} For $ c \in \mathbb{R^+}$ and  $\alpha \in \mathbb{R}$ define a dilation matrix $D_c $, with $c$ as the dilation parameter and boost matrix $B_\alpha$, with $ \alpha$ as boost parameter operating on a vector in space-time as
	$\mu =(s,t)^T \in \mathbb{R}^2 $ as
	$$  D_c=\begin{pmatrix}c&0\\0&c\end{pmatrix} ~~ and ~~ B_\alpha=\begin{pmatrix}\cosh\alpha&-\sinh\alpha\\-\sinh\alpha&\cosh\alpha
	\end{pmatrix}.  $$
	The composition of above two transformations produce a boost-dilation matrix $ M_{c,\alpha}$ defined by
	$$ M_{c,\alpha}=
	\\		  \begin{pmatrix}	c\cosh\alpha&-c\sinh\alpha\\-c\sinh\alpha&c\cosh\alpha
	\end{pmatrix}.\\$$ 
	Now the family of boostlets w.r.t the mother boostlet $\varphi\in L^2(\mathbb{R}^2)$ is defined as
	$$\varphi_{c,\alpha,\uptau}(\mu)=c^{-1}  \varphi(M_{c,\alpha}^{-1}(\mu-\uptau)),$$
	where $ \uptau=(\uptau_s,\uptau_t)^T \in \mathbb{R}^2$ is a translation vector in 2D space-time.
	
	\parindent=0mm \vspace{0.1in}
	
	\textbf{Definition 2.2.}
	Define a set $ \mathbb{S}=\{(c,\alpha,\uptau):c\in\mathbb{R}^+,\alpha\in\mathbb{R},\uptau\in\mathbb{R}^2\} $, with product $\cdot$ such that 
	$$ (c,\alpha,\uptau)\cdot (c',\alpha',\uptau') =(cc',\alpha + \alpha' ,\uptau+\textbf{B}_\alpha \textbf{D}_c\uptau') $$
	This set forms a group known as boostlet group, with (1,0,0) as identity of the group. Now
	for a near field mother boostlet $ \varphi \in L^2(\mathbb{R}^2)$, we define the continuous boostlet transform ~\cite{emj} of a function $ f(\mu)\in L^2(\mathbb{R}^2)$ involving space-time coordinates as follows :
	\begin{equation}
		\textbf{B}_\varphi f(c,\alpha,\uptau)= \left( \langle f,\varphi_{c,\alpha,\uptau}\rangle,\langle f,
		\varphi^{*}_{c,\alpha,\uptau}\rangle\right)_{(c,\alpha,\uptau) \in \mathbb{S}}.\label{1} 
	\end{equation}
	
	\parindent=0mm \vspace{0.1in}
	\textbf{Definition 2.3.}
	A window function $ \varphi \in L^2(\mathbb{R}^2) $  is said to be admissible in 2D space-time if $ \Delta $ defined by
	\begin{eqnarray*}
		\Delta&=&\int_{\mathbb{R}} \int_{0}^{\infty} \big|\hat{\varphi}(M
		^{T}_{c,\alpha}\omega)\big|^{2} \frac{dc d\alpha}{c} + \int_{\mathbb{R}} \int_{0}^{\infty} \big|\hat{\varphi}^{*}(M^{T}
	_{c,\alpha}\omega)\big|^{2} \frac{dc d\alpha}{c}\\\\&=&\Delta_{\varphi}(\omega)+\Delta_{\varphi^{*}}(\omega)
	\end{eqnarray*}
	is a constant independent of $ \omega $ satisfying $ 0 < \Delta < \infty $.

	\parindent=0mm \vspace{.1in}
	{\bf{Proposition 2.1.}} Given an admissible boostlet $\varphi \in L^2(\mathbb{R}^2) $, the continuous boostlet transform (1) of an arbitrary function $ f\in L^{2}(\mathbb{R}^2) $, can be expressed as
	\begin{equation}
		\textbf{B}_\varphi f(c,\alpha,\uptau)=\left( \left( f*\check{\varphi^{*}}_{c,\alpha,0}\right), \left( f*\check{\varphi}_{c,\alpha,0} \right)  \right)     
	\end{equation}
	
	\parindent=0mm \vspace{.1in}
	\textbf{Proof}. From the definition (1), we have 
	\begin{equation*}
		\textbf{B}_\varphi f(c,\alpha,\uptau)=\left( \langle f,\varphi_{c,\alpha,\uptau}\rangle,\langle f,\varphi^{*}_{c,\alpha,\uptau}\rangle \right).\end{equation*}
	Now, we have 
	\begin{eqnarray*}
		\langle f,\varphi_{c,\alpha,\uptau} \rangle &=&\int_{\mathbb{R}^2}f(\mu)\varphi^{*}_{c,\alpha,\uptau}(\mu)d\mu\\\\&=&\int_{\mathbb{R}^2}f(\mu)c^{-1}\varphi^{*}(M^{-1}_{c,\alpha}(\mu-\uptau))d\mu\\\\&=&\int_{\mathbb{R}^2}c^{-1}f(\mu)\check{\varphi ^{*}}(M^{-1}_{c,\alpha}(\uptau-\mu))d\mu\\\\&=&\left( f*\check{\varphi ^{*}}_{c,\alpha,0} \right)(\uptau).
	\end{eqnarray*}
	where $\check{\varphi}(\mu)=\varphi(-\mu)$.
	
	Also,
	\begin{eqnarray*}
		\langle f,\varphi^{*}_{c,\alpha,\uptau}\rangle&=&\int_{\mathbb{R}^2}f(\mu)\varphi_{c,\alpha,\uptau}d\mu\\\\&=&\int_{\mathbb{R}^2}c^{-1}f(\mu)\varphi {M_{c,\alpha}^{-1}(\mu-\uptau)}d\mu\\\\&=&\int_{\mathbb{R}^2}c^{-1}f(\mu)\check{ \varphi}M_{c,\alpha}^{-1}(\uptau-\mu)d\mu\\\\&=&\left(f*\check\varphi_{c,\alpha,0}\right)(\uptau),
	\end{eqnarray*}
	where $\check{\varphi}(\mu)=\varphi(-\mu).$
	
	Therefore, we have
	$$\textbf{B}_\varphi f(c,\alpha,\uptau)=\left( \langle f\;,\;\varphi_{c,\alpha,\uptau}\rangle,\langle f\;,\;\varphi^{*}_{c,\alpha,\uptau}\rangle\right)=\left(	\left(f*\check\varphi^{*}_{c,\alpha,0}\right)(\uptau),\left(f*\check\varphi_{c,\alpha,0}\right)(\uptau) \right). $$
	which  completes the proof.
	
	\parindent=8mm \vspace{0.1in}
	
	The subsequent theorem rigorously states and proves the fundamental properties of the continuous boostlet transform.
	
	\parindent=0mm \vspace{0.1in}
	{\bf{Theorem 2.2.}}
	For the boostlets $ \varphi$ and $\psi$ and the arbitrary functions $f_1 $ and $ f_2 $ in $L^{2}(\mathbb{R}^2)$, the subsequent properties hold for the boostlet transform:
	\begin{enumerate}
		\item \textbf {Linearity} : $ \textbf{B}_\varphi(\ell  f_1 + m   f_2)(c,\alpha,\uptau)$ = $\ell  \textbf{B}_\varphi f_1(c,\alpha,\uptau)+m \textbf{B}_\varphi f_2(c,\alpha,\uptau),$ where $ \ell,m \in \mathbb {C}$.
		
		\parindent=0mm \vspace{.2in}
		\textbf{Proof.} We have
		\begin{eqnarray*}
			\textbf{B}_\varphi(\ell f_1 +m f_2)(c,\alpha,\uptau)&=&\left( \langle \ell  f_1+m f_2,\varphi_{c,\alpha,\uptau} \rangle,\langle \ell f_1+ m f_2,\varphi^{*}_{c,\alpha,\uptau} \rangle \right)\\\\&=&\left(\ell \langle f_1,\varphi_{c,\alpha,\uptau} \rangle + m \langle f_2,\varphi_{c,\alpha,\uptau} \rangle,\ell  \langle f_1,\varphi^{*}_{c,\alpha,\uptau} \rangle +m \langle f_2,\varphi^{*}_{c,\alpha,\uptau}\rangle \right)\\\\&=&\ell \left( \langle f_1,\varphi_{c,\alpha,\uptau} \rangle ,\langle f_1,\varphi^{*}_{c,\alpha,\uptau} \rangle \right)+ m \left( \langle f_2,\varphi_{c,\alpha,\uptau} \rangle ,\langle f_2,\varphi^{*}_{c,\alpha,\uptau} \rangle \right)\\\\&=&\ell  \textbf{B}_\varphi f_1(c,\alpha,\uptau)+m \textbf{B}_\varphi f_2(c,\alpha,\uptau).~~~~\square
		\end{eqnarray*}
		
		\item \textbf{Quadratic Homogeneity}: $\textbf{B}_{a \varphi+b \psi} f(c,\alpha,\uptau) = a a^{*} \textbf{B}_\varphi f(c,\alpha,\uptau)+b b^{*}\textbf{B}_\psi f(c,\alpha,\uptau),$ where $ a,b \in \mathbb{C} $.
		
		\parindent=0mm \vspace{.1in}
		\textbf{Proof.} We have
		\begin{eqnarray*}
			\textbf{B}_{a \varphi +b \psi}f (c,\alpha ,\uptau)&=&\left( \langle  f,a \varphi_{c,\alpha,\uptau}+b \psi_{c,\alpha,\uptau} \rangle,\langle f,{(a\varphi_{c,\alpha,\uptau}+b \psi_{c,\alpha,\uptau})}^{*}\rangle \right)\\\\&=& \left( a^{*} \langle f, \varphi_{c,\alpha,\uptau} \rangle +b^{*} \langle f,\psi_{c,\alpha,\uptau} \rangle,a \langle f,\varphi^{*}_{c,\alpha,\uptau} \rangle +b \langle f,\psi^{*}_{c,\alpha,\uptau} \rangle \right)\\\\&=&\left( a^{*}\langle f,\varphi_{c,\alpha,\uptau}\rangle,a \langle f,\varphi_{c,\alpha,\uptau}^{*} \rangle \right) +\left( b^{*} \langle f,\psi_{c,\alpha,\uptau} \rangle ,b \langle f,\psi^{*}_{c,\alpha,\uptau} \rangle \right)\\\\&=&a a^{*} \left( \langle f,\varphi_{c,\alpha,\uptau} \rangle ,\langle f,\varphi^{*}_{c,\alpha,\uptau} \rangle \right) +b b^{*} \left( \langle f,\psi_{c,\alpha,\uptau} \rangle,\langle f,\psi_{c,\alpha,\uptau}^{*} \rangle \right) \\\\&=& a a^{*} \textbf{B}_\varphi f(c,\alpha,\uptau)+b b^{*}\textbf{B}_\psi f(c,\alpha,\uptau).~~~~~\square
		\end{eqnarray*}

		\item \textbf{Translation} : $ \textbf{B}_\varphi T_k f(c,\alpha,\uptau) = \textbf{B}_\varphi f(c,\alpha ,\uptau -k) $ : where $T_k$ is the translation operator defined by $ T_k f(\mu)=f(\mu-k). $
		
		\parindent=0mm \vspace{.1in}
		\textbf{Proof.} We have
		\begin{equation}	\textbf{B}_\varphi T_k f(c,\alpha,\uptau) = \left( \langle T_k f,\varphi_{c,\alpha,\uptau} \rangle ,\langle T_k f,\varphi^{*}_{c,\alpha,\uptau} \rangle \right).
		\end{equation}
		Now
		\begin{eqnarray*}
			\langle T_k f,\varphi_{c,\alpha,\uptau} \rangle &=&\int_{\mathbb{R}^2} T_k f(\mu) \varphi_{c,\alpha,\uptau}^{*} d\mu\\\\&=&\int_{\mathbb{R}^2}f(\mu-k)\varphi_{c,\alpha,\uptau}^{*}(\mu)d\mu\\\\&=&\int_{\mathbb{R}^2}f(\mu-k)c^{-1}\varphi^{*}(M_{c,\alpha}^{-1}(\mu-k))d\mu\\\\&=&c^{-1}\int_{\mathbb{R}^2}f(y)\varphi^{*}(M_{c,\alpha}^{-1}(k+y-\uptau))dy\\&=&c^{-1}\int_{\mathbb{R}^2}f(y)\varphi^{*}(M_{c,\alpha}^{-1}(y-(\uptau-k)))dy\\\\&=&\int_{\mathbb{R}^2}f(y)\varphi^{*}_{c,\alpha,\uptau-k}dy\\\\&=&\langle f,\varphi_{c,\alpha,\uptau-k} \rangle.
		\end{eqnarray*} 
		Also,
		\begin{eqnarray*}
			\langle T_k f,\varphi^{*}_{c,\alpha,\uptau} \rangle &=&\int_{\mathbb{R}^2}T_k f(\mu)\varphi_{c,\alpha,\uptau}d\mu\\&=&\int_{\mathbb{R}^2}f(\mu-k)c^{-1}\varphi(M^{-1}_{c,\alpha}(\mu-\uptau))d\mu\\\\&=&\int_{\mathbb{R}^2}f(y)c^{-1}\varphi(M^{-1}_{c,\alpha}(y+k-\uptau))dz\\&=&\int_{\mathbb{R}^2}f(y)c^{-1}\varphi(M^{-1}_{c,\alpha}(y-(k-\uptau)))dy\\\\&=&\int_{\mathbb{R}^2}f(y)\varphi_{c,\alpha,\uptau-k}(y)dy\\\\&=&\langle f,\varphi^{*}_{c,\alpha,\uptau-k} \rangle.
		\end{eqnarray*}
		Therefore, we have 
		\begin{eqnarray*}
			\textbf{B}_\varphi T_k f(c,\alpha,\uptau)&=&\left( \langle\varphi_{c,\alpha,\uptau-k} \rangle,\langle f,\varphi^{*}_{c,\alpha,\uptau-k} \rangle \right)\\\\&=&\textbf{B}_\varphi f(c,\alpha,\uptau-k).~~\square
		\end{eqnarray*}
		\item \textbf{Scaling}: $\textbf{B}_\varphi f(\lambda \mu) (c,\alpha,\uptau)=\frac{1}{\lambda} \textbf{B}_{\varphi'} f(c,\alpha,\uptau \lambda):$
		 where $\varphi'(\mu)=\varphi\left(\frac{\mu}{\lambda}\right) $
		 and $ \lambda \in \mathbb{R^{+}}.$
		
		\parindent=0mm \vspace{.1in}
		\textbf{Proof}. 
		\begin{equation*}
			\textbf{B}_\varphi f(\lambda \mu)(c,\alpha,\uptau)=\left( \langle f(\lambda \mu),\varphi_{c,\alpha,\uptau} \rangle ,\langle f(\lambda \mu),\varphi^{*}_{c,\alpha,\uptau} \rangle \right)
		\end{equation*}
		Now, 
		\begin{eqnarray*}
			\left\langle f(\lambda \mu),\varphi_{c,\alpha,\uptau} \right\rangle&=&\int_{\mathbb{R}^2}f(\lambda \mu) \varphi_{c,\alpha,\uptau}d\mu\\\\&=&\int_{\mathbb{R}^2}f(\lambda \mu)c^{-1}\varphi^{*}(M^{-1}_{c,\alpha}(\mu-\uptau))d\mu\\\\&=&\frac{c^{-1}}{\lambda}\int_{\mathbb{R}^2}f(y)\varphi^{*}(M^{-1}_{c,\alpha}(\frac{y}{\lambda}-\uptau))dy\\\\&=&\frac{1}{\lambda}\int_{\mathbb{R}^2}f(y)c^{-1}\varphi'^{*}(M^{-1}_{c,\alpha}(y-\lambda \uptau))dy\\\\&=&\frac{1}{\lambda}\langle f,\varphi'_{c,\alpha,\lambda\uptau} \rangle ; where ~~~~~  \varphi'(\mu) = \varphi\left(\frac{\mu}{\lambda}\right) .
		\end{eqnarray*}
		Also,
		\begin{eqnarray*}
			\langle f(\lambda \mu),\varphi^{*}_{c,\alpha,\uptau} \rangle&=&\int_{\mathbb{R}^2}f(\lambda\mu)\varphi_{c,\alpha,\uptau}(\mu)d\mu\\\\&=&\int_{\mathbb{R}^2}f(\lambda\mu)c^{-1}\varphi(M^{-1}_{c,\alpha}(\mu-\uptau))d\mu\\\\&=&\frac{c^{-1}}{\lambda}\int_{\mathbb{R}^2}f(y)\varphi(M^{-1}_{c,\alpha}(\frac{y}{\lambda}-\uptau))dy\\\\&=&\frac{1}{\lambda}\int_{\mathbb{R}^2}f(y)\varphi'(M^{-1}_{c,\alpha}(y-\lambda\uptau))dy\\\\&=&\frac{1}{\lambda}\langle f,\varphi'^{*}_{c,\alpha,\lambda\uptau} \rangle,\textit{   where ~}~ \varphi'(\mu) = \varphi\left(\frac{\mu}{\lambda}\right).
		\end{eqnarray*}
		Therefore, we have 
		\begin{eqnarray*}
			\textbf{B}_\varphi f(\lambda \mu)(c,\alpha,\uptau)&=&\left( \frac{1}{\lambda}\langle f,\varphi'_{c,\alpha,\lambda\uptau} \rangle,\frac{1}{\lambda}\langle f,\varphi'^{*}_{c,\alpha,\lambda\uptau} \rangle \right)\\\\&=&\frac{1}{\lambda} \textbf{B}_{\varphi'} f(c,\alpha,\lambda\uptau), 
		\end{eqnarray*}
		  where~$\varphi'(\mu)=\varphi\left(\frac{\mu}{\lambda}\right).~~~\square$
		\item \textbf{Reflection}: $\textbf{B}_\varphi f(-\mu)(c,\alpha,\uptau) =-\textbf{B}_{\check{\varphi}} f(c,\alpha,-\uptau)$
		\parindent=0mm \vspace{.1in}
		
		\textbf{Proof.}
		
		\begin{equation*}
			\textbf{B}_\varphi f(-\mu)(c,\alpha,\uptau)=\left( \langle f(-\mu) ,\varphi_{c,\alpha,\uptau}(\mu)\rangle,\langle f(-\mu),\varphi^{*}_{c,\alpha,\uptau}(\mu) \rangle \right)
		\end{equation*}
		Now
		\begin{eqnarray*}
			\langle f(-\mu) ,\varphi_{c,\alpha,\uptau}(\mu)\rangle&=&\int_{\mathbb{R}^2}f(-\mu)\varphi^{*}_{c,\alpha,\uptau}d\mu\\\\&=&\int_{\mathbb{R}^2}f(-\mu)c^{-1}\varphi^{*}(M^{-1}_{c,\alpha}(\mu-\uptau))d\mu\\\\&=&-\int_{\mathbb{R}^2}f(y)c^{-1}\varphi^{*}(M^{-1}_{c,\alpha}(-y-\uptau))dy\\\\&=&-\int_{\mathbb{R}^2}f(y)\check\varphi^{*}_{c,\alpha,-\uptau}(y)dy\\&=&-\langle f,\check \varphi _{c,\alpha,-\uptau} \rangle,~\textit{where}~~~\check \varphi(\mu)=\varphi(-\mu).
		\end{eqnarray*}
		Also,
		\begin{eqnarray*}
			\langle f(-\mu),\varphi^{*}_{c,\alpha,\uptau} \rangle&=&\int_{\mathbb{R}^2}f(-\mu) \varphi_{c,\alpha,\uptau}d\mu\\\\&=&\int_{\mathbb{R}^2} f(-\mu)c^{-1}\varphi(M^{-1}_{c,\alpha}(\mu-\uptau))d\mu\\\\&=&-\int_{\mathbb{R}^2}f(y)c^{-1}\check \varphi(M^{-1}_{c,\alpha}(y+\uptau))dy\\\\&=&-\langle f,\check{\varphi^{*}}_{c,\alpha,-\uptau} \rangle.
		\end{eqnarray*}
		Hence,
		\begin{eqnarray*}
			\textbf{B}_\varphi f(-\mu)(c,\alpha,\uptau)&=&\left( -\langle f,\check{\varphi}_{c,\alpha,-\uptau} \rangle,-\langle f,\check{\varphi^{*}}_{c,\alpha,-\uptau} \rangle \right)\\\\&=&-\textbf{B}_{\check{\varphi}} f(c,\alpha,-\uptau). ~~~~~~~~~\square
		\end{eqnarray*}   
	\end{enumerate}

	\parindent=0mm \vspace{0.1in}
	{\bf{Theorem 2.3.}}
	Let $ \textbf{B}_\varphi f(c,\alpha,\uptau) $ denotes the boostlet transform of an arbitrary function $ f \in L^2(\mathbb{R}^{2})$, then
	\begin{equation}
		\mathfrak{F}\{ \textbf{B}_\varphi f(c,\alpha,\uptau)\}(\omega)=\left( c \hat{f}(\omega){\hat{\varphi}}^{*}(M^{T}_{c,\alpha}\omega),c\hat{f}(\omega)\check{\hat{\varphi}}(M^{T}_{c,\alpha}\omega) \right),
	\end{equation} 
	where $\mathfrak{F}$ denotes the Fourier transform of spatio-temporal function $f(\mu)$ given by
	\begin{equation*}
	\mathfrak{F}(f)(\omega)=	\hat{f}(\omega)=\int_{\mathbb{R}^2}f(\mu)e^{-2\pi \iota\omega^{T}\mu}d\mu
	\end{equation*}
	
	\parindent=0mm \vspace{.1in}
	\textbf{Proof}. 
	For any function $ f \in L^2(\mathbb{R}^2) $ ,we have
	\begin{equation*}
		\mathfrak{F} \{ \textbf{B}_\varphi f(c,\alpha,\uptau) \}(\omega)=\left( \mathfrak{F}\langle f,\varphi_{c,\alpha,\uptau} \rangle,\mathfrak{F}\langle f,\varphi^{*}_{c,\alpha,\uptau} \rangle \right).
	\end{equation*}
	The Fourier transform of $\varphi_{c,\alpha,\uptau}$ is given by
	\begin{equation*}
		\hat\varphi_{c,\alpha,\uptau}(\omega)=c \hat \varphi (M^{T}_{c,\alpha}\omega)e^{-2\pi\iota\uptau^{T}\omega}.
	\end{equation*}
	We have
	\begin{eqnarray*}
		\langle f,\varphi_{c,\alpha,\uptau}\rangle&=&\langle \hat{f}, \hat \varphi_{c,\alpha,\uptau}\rangle\\&=&\int_{\mathbb R^2} \hat{f}(\omega)\hat \varphi^{*}_{c,\alpha,\uptau}(\omega)d\omega\\&=&\int_{\mathbb R^2}\hat{f}(\omega)  c e^{2\pi\iota\uptau^{T}\omega} \hat\varphi^{*} (M^{T}_{c,\alpha}\omega) d\omega\\&=&\mathfrak{F}^{-1}\{ c\hat{f}(\omega) \hat{\varphi}^{*}(M^{T}_{c,\alpha}\omega)\}\\
		\mathfrak{F}\langle f,\varphi_{c,\alpha,\uptau} \rangle& =& c \hat{f}(\omega)\hat{\varphi}^{*}(M^{T}_{c,\alpha}\omega). 
	\end{eqnarray*}
	Also
	\begin{eqnarray*}
		\langle f,\varphi^{*}_{c,\alpha,\uptau} \rangle&=&\langle \hat{f}(\omega) ,\widehat{\varphi^{*}_{c,\alpha,\uptau}}(\omega) \rangle\\&=&\langle \hat{f}(\omega),{\hat\varphi}^{*}_{a,\alpha,\uptau}(-\omega)\rangle\\&=&\int_{\mathbb R^2}\hat{f}(\omega)\hat \varphi_{c,\alpha,\uptau}(-\omega)d\omega\\&=&\int_{\mathbb R^2}\hat{f}(\omega)c e^{2\pi\iota\uptau^{T}\omega}\hat{\varphi}(-M^{T}_{c,\alpha}\omega)d\omega\\&=&\int_{\mathbb R^2}\hat{f}(\omega) c\check{\hat{\varphi}}(M^{T}_{c,\alpha}\omega) \omega)e^{2\pi\iota\uptau^{T}\omega} d\omega\\&=&\mathfrak{F}^{-1}\{c\hat{f}(\omega)\check{\hat{\varphi}}(M^{T}_{c,\alpha}\omega)\}\\\mathfrak{F}\langle f,\varphi^{*}_{c,\alpha,\uptau} \rangle&=&c\hat{f}(\omega)\check{\hat{\varphi}}(M^{T}_{c,\alpha}\omega), ~~~where~ \check{ \varphi}(\mu)=\varphi(-\mu).
	\end{eqnarray*}
	Hence,
	\begin{equation*}
		\mathfrak{F}\{ \textbf{B}_\varphi f(c,\alpha,\uptau)\}(\omega)=\left( c\hat{f}(\omega){\hat{\varphi}}^{*}(M^{T}_{c,\alpha}\omega),c \hat{f}(\omega)\check{\hat{\varphi}}(M^{T}_{c,\alpha}\omega) \right).~~\square
	\end{equation*}
	
	{\bf{3. Uncertainty Principles Associated with Boostlet Transform}}
	
	\parindent=0mm \vspace{0.1in}   
	The uncertainty principle also known as duration-bandwidth principle is an elementary principle of harmonic analysis, which states that a signal which is very concentrated in time has its Fourier transform outspread and vice versa. In otherwords an arbitrary function cannot be compact both in time and frequency~\cite{ga,gjg,sv,pl}. In this section we shall present some uncertainty principles including Heisenberg's uncertainty principle, Logarithmic, Pitt's and Nazarov's uncertainty principles for the boostlet transform.

	\parindent=0mm \vspace{0.1in}
	{\bf{Theorem 3.1.}} If $\textbf{B}_\varphi \psi(c,\alpha,\uptau)$ denotes the boostlet transform of any non trivial function $\psi \in L^2(\mathbb{R}^2 ) $ with respect to the admissible boostlet $ \varphi \in L^2(\mathbb{R}^2)$ then the following uncertainty inequality holds:
	\begin{equation}
		\Biggl\{ \int_{\mathbb{S}}|\uptau|^2 |\textbf{B}_\varphi \psi(c,\alpha,\uptau)|^2\frac{dc d\alpha d\uptau}{c^{3}} \Biggr\}^{\frac{1}{2}} \Biggl\{\int_{\mathbb{R}^2} |\gamma|^2 |\hat{\psi}(\gamma)|^2 d\gamma \Biggr\}^{\frac{1}{2}} \geq\frac{1}{2}\sqrt{\Delta}||\psi||^2.
	\end{equation}
	
	\textbf{Proof.} The classical Heisenberg-Pauli-Weyl inequality in time and frequency domain for any non trivial function $\psi \in L^2(\mathbb{R}^2)$, is given by~\cite{cp}
	\begin{align*}
		\Biggl\{ \int_{\mathbb{R}^2}|t|^2 |\psi(t)|^2dt \Biggr\}^{\frac{1}{2}} \Biggl\{ \int_{\mathbb{R}^2}|\gamma|^2 |\hat{\psi}(\gamma)|^2d\omega \Biggr\}^{\frac{1}{2}} \geq\frac{1}{2} \int_{\mathbb{R}^2}|f(t)|^2 dt.
	\end{align*}
	Since $ \textbf{B}_\varphi \psi(c,\alpha,\uptau) \in L^2(\mathbb{R}^2)$ whenever $ \psi \in L^2(\mathbb{R}^2) $,therefore replacing $ \psi$ by $ \textbf{B}_\varphi \psi(c,\alpha,\uptau)$ in above equality to obtain
	\begin{align*}
		\Biggl\{ \int_{\mathbb{R}^2}|\uptau|^2 |\textbf{B}_\varphi \psi(c,\alpha,\uptau)|^2 d\uptau \Biggr\}^{\frac{1}{2}} \Biggl\{ \int_{\mathbb{R}^2}|\gamma|^2 |\mathfrak{F} \{\textbf{B}_\varphi \psi(c,\alpha,\uptau) \}(\gamma)|^2d\gamma \Biggr\}^{\frac{1}{2}} \geq\frac{1}{2} \int_{\mathbb{R}^2}|\textbf{B}_\varphi \psi(c,\alpha,\uptau)|^2 d \uptau.
	\end{align*}
	Now integrate the above inequality w.r.t the measure $\dfrac{dc d\alpha}{c^{3}}$ so that
	\begin{align*}
		\int_{\mathbb{R}} \int_\mathbb{R^{+}}{\Biggl\{\Biggl\{ \int_{\mathbb{R}^2}|\uptau|^2 |\textbf{B}_\varphi \psi(c,\alpha,\uptau)|^2 d\uptau \Biggr\}^{\frac{1}{2}} \Biggl\{ \int_{\mathbb{R}^2}|\gamma|^2 |\mathfrak{F}
			\{\textbf{B}_\varphi \psi(c,\alpha,\uptau) (\gamma)|^2d\gamma \Biggr\}^{\frac{1}{2}}\Biggr\}} \frac{dc d\alpha}{c^{3}}\\  \geq\frac{1}{2} \int_{\mathbb{R}}\int_{\mathbb{R}^{+}} \Biggl\{  \int_{\mathbb{R}^2}|\textbf{B}_\varphi \psi(c,\alpha,\uptau)|^2 d\uptau \Biggr\} \frac{dc d\alpha}{c^{3}}. 
	\end{align*}
	Using Cauchy Schwartz inequality and Fubini's theorem, we have
	\begin{align*}
		\Biggl\{ \int_{\mathbb{R}^2 \times \mathbb{R} \times {\mathbb{R}^{+}}} |\uptau|^2 |\textbf{B}_\varphi \psi(c,\alpha,\uptau)|^2 \frac{dc d\alpha d\uptau}{c^{3}} \Biggr\}^{\frac{1}{2}} \Biggl\{ \int_{\mathbb{R}^2 \times \mathbb{R} \times \mathbb{R}^{+}}|\gamma|^2 |\mathfrak{F} \{\textbf{B}_\varphi \psi(c,\alpha,\uptau)\} (\gamma)|^2\frac{dc d\alpha d\gamma}{c^{3}} \Biggr\}^{\frac{1}{2}}\\  \geq\frac{1}{2} \int_{\mathbb{R}^2  \times \mathbb{R} \times \mathbb{R}^+}  |\textbf{B}_\varphi \psi(c,\alpha,\uptau)|^2 \frac{dc d\alpha d\uptau}{c^{3}} . 
	\end{align*}
	\begin{align*}
		\Biggl\{ \int_{\mathbb{R}^2 \times \mathbb{R}  \times \mathbb{R}^{+}} |\uptau|^2 |\textbf{B}_\varphi \psi(c,\alpha,\uptau)|^2 \frac{dc d\alpha d\uptau}{c^{3}} \Biggr\}^{\frac{1}{2}} \Biggl\{ \int_{\mathbb{R}^2 \times \mathbb{R} \times \mathbb{R}^+} |\gamma|^2 |\mathfrak{F} \{\textbf{B}_\varphi \psi(c,\alpha,\uptau)\} (\gamma)|^2\frac{dc d\alpha d\gamma}{c^{3}} \Biggr\}^{\frac{1}{2}}\\  \geq\frac{1}{2} \Delta ||\psi||^2 . 
	\end{align*}
	Evaluation of the second integral on the left hand side of above inequality yields
	\begin{eqnarray*}
		&&\displaystyle \int_{\mathbb{R}^2 \times \mathbb{R} \times \mathbb{R}^+}|\gamma|^2 |\mathfrak{F} \{\textbf{B}_\varphi \psi(c,\alpha,\uptau)\} (\gamma)|^2\frac{dc d\alpha d\gamma}{c^{3}}\\\\
		&&\qquad\qquad=\int_{\mathbb{R}^2 \times \mathbb{R} \times \mathbb{R}^+}|\gamma|^2\Bigl\{ |c|^2|\hat{\psi}(\gamma) |^2 |{\hat{\varphi}}^{*}(M^{T}_{c,\alpha}\gamma)|^2+ |c|^2 |\hat{\psi}(\gamma)|^2 |\check{{\hat{\varphi}}}(M^{T}_{c,\alpha}\gamma)|^2  \Bigr\}  \frac{dc d\alpha d\gamma}{c^{3}}\\\\
		&&\qquad\qquad=  \int_{\mathbb{R}^2  \times \mathbb{R} \times \mathbb{R}^+} |\gamma|^2 |c|^2 |\hat{\psi}(\gamma)|^2 \{ |{\hat{\varphi}}^{*} (M^{T}_{c,\alpha}\gamma)|^2  +|\check{\hat{\varphi}}(M^{T}_{c,\alpha}(\gamma))|^2  \} \frac{dc d\alpha d\gamma}{c^{3}}\\\\
		&&\qquad\qquad= \int_{\mathbb{R}^2}|\gamma|^2 |\hat{\psi}(\gamma)|^2 d\gamma \int_{\mathbb{R} \times \mathbb{R}^+} \{ |{\hat{\varphi}}^{*} (M^{T}_{c,\alpha}\gamma)|^2  +|\check{\hat{\varphi}}(M^{T}_{c,\alpha}(\gamma))|^2 |\} \frac{dc d\alpha}{c}\\\\
		&&\qquad\qquad=\int_{\mathbb{R}^2}|\gamma|^2 |\hat{\psi}(\gamma)|^2 d\gamma \times [ \Delta_\varphi^{*} +\Delta_\varphi ] \\\\
		&&\qquad\qquad=  \int_{\mathbb{R}^2}|\gamma|^2 |\hat{\psi}(\gamma)|^2 d\gamma  \times \Delta.
	\end{eqnarray*} 
	Substituting this in to above inequality, we get
	\begin{align*}
		\Biggl\{ \int_{\mathbb{R}^2 \times \mathbb{R} \times \mathbb{R}^{+}} |\uptau|^2 |\textbf{B}_\varphi \psi(c,\alpha,\uptau)|^2 \frac{dc d\alpha d\uptau}{c^{3}} \Biggr\}^{\frac{1}{2}}\Biggl\{ \int_{\mathbb{R}^2}|\gamma|^2 |\hat{\psi}(\gamma)|^2 d\gamma \Biggr\}^\frac{1}{2} \times {\Delta}^\frac{1}{2}\geq \frac{1}{2} \Delta ||\psi||^2,
	\end{align*}
	which implies,
	\begin{align*}
		\Biggl\{ \int_{\mathbb{S}} |\uptau|^2 |\textbf{B}_\varphi \psi(c,\alpha,\uptau)|^2 \frac{dc d\alpha d\uptau}{c^{3}} \Biggr\}^{\frac{1}{2}}\Biggl\{ \int_{\mathbb{R}^2}|\gamma|^2 |\hat{\psi}(\gamma)|^2 d\gamma \Biggr\}^\frac{1}{2} \geq \frac{1}{2} \sqrt{\Delta} ||\psi||^2.~\square
	\end{align*} 
	
	\parindent=8mm \vspace{0.1in}
	In the subsequent theorem, we intend to establish the generalization of theorem 3.1 within the space $ L^p(\mathbb{R}^2)$, for  $1 \leq p \leq2 $ and $p \geq 2 $ given by:
	
	\parindent=0mm \vspace{0.1in}
	{\bf{Theorem 3.2.}}
	Given an admissible boostlet function $ \varphi \in L^2(\mathbb{R}^2) $ and a non-trivial function $ \psi \in  L^2(\mathbb{R}^2) $,  we have
	
	\parindent=0mm \vspace{0.1in}  
	(i)  \begin{align}
		\Biggl\{ \int_{\mathbb{S}}  |\uptau \textbf{B}_\varphi \psi(c,\alpha,\uptau)|^{p}\frac{dc d\alpha d\uptau}{c^{3}} \Biggr\}^{\frac{1}{p}} \Biggl\{\int_{\mathbb{R}^2} |\gamma \hat{\psi}(\gamma)|^{p} d\gamma \Biggr\} ^{\frac{1}{p}} \geq\frac{\sqrt{\Delta}}{2}||\psi||^2, ~\textit{for}~ 1 \leq p \leq 2.
	\end{align}
	
	\parindent=0mm \vspace{0.1in} 			
	(ii) \begin{align}
		\Biggl\{ \int_{\mathbb{S}}  |\uptau|^{p} |\textbf{B}_\varphi \psi(c,\alpha,\uptau)|^{2}\frac{dc d\alpha d\uptau}{c^{3}} \Biggr\}^{\frac{1}{p}} \Biggl\{\int_{\mathbb{R}^2} |\gamma|^{p} |\hat{\psi}(\gamma)|^{2} d\gamma \Biggr\} ^{\frac{1}{p}} \geq {\Delta}^{\frac{1}{p}}||\psi||^{\frac{4}{p}},  ~\textit{for}~   p\geq 2   	  
	\end{align}
	
	\parindent=0mm \vspace{.1in}
	\textbf{Proof}  (i)   For any non-trivial function $ \psi \in L^2(\mathbb{R}^2) $ the dispersion in time and frequency satisfies the inequality~\cite{cp}
	\begin{align*}
		\Biggl\{ \int_{\mathbb{R}^2}|t \psi(t)|^{p}dt \Biggr\}^{\frac{1}{p}} \Biggl\{ \int_{\mathbb{R}^2}|\gamma \hat{\psi}(\gamma)|^{p}d\gamma \Biggr\}^{\frac{1}{p}} \geq\frac{1}{2} \int_{\mathbb{R}^2}|\psi(t)|^2 dt.
	\end{align*}
	Since $ \textbf{B}_\varphi \psi(c,\alpha,\uptau) \in L^2(\mathbb{R}^2)$ whenever $ \psi \in L^2(\mathbb{R}^2) $, therefore replacing $ \psi$ by $ \textbf{B}_\varphi \psi(c,\alpha,\uptau)$ in above equality to obtain
	\begin{align*}
		\Biggl\{ \int_{\mathbb{R}^2}|\uptau\textbf{B}_\varphi \psi(c,\alpha,\uptau)|^{p} d\uptau \Biggr\}^{\frac{1}{p}} \Biggl\{ \int_{\mathbb{R}^2}|\gamma \mathfrak{F} \{\textbf{B}_\varphi \psi(c,\alpha,\uptau) \}(\gamma)|^{p}d\gamma \Biggr\}^{\frac{1}{p}} \geq\frac{1}{2} \int_{\mathbb{R}^2}|\textbf{B}_\varphi \psi(c,\alpha,\uptau)|^2 d\uptau.
	\end{align*}
	Integrating w.r.t measure $ \dfrac{dc d\alpha}{c^{3}} $ so that
	\begin{eqnarray*}
		\int_{\mathbb{R}} \int_{\mathbb{R}^{+}} \Biggl\{ \Biggl\{ \int_{\mathbb{R}^2}|\uptau\textbf{B}_\varphi \psi(c,\alpha,\uptau)|^{p} d\uptau \Biggr\}^{\frac{1}{p}} \Biggl\{ \int_{\mathbb{R}^2}|\gamma \mathfrak{F} \{\textbf{B}_\varphi \psi(c,\alpha,\uptau) \}(\gamma)|^{p}d\gamma \Biggr\}^{\frac{1}{p}} \Biggr\}  \dfrac{dc d\alpha}{c^{3}}\\\\
		\qquad\qquad	 \geq\frac{1}{2}\int_{\mathbb{R}} \int_{\mathbb{R}^{+}} \Biggl\{ \int_{\mathbb{R}^2}|\textbf{B}_\varphi \psi(c,\alpha,\uptau)|^2 d\uptau\Biggr\} \dfrac{dc d\alpha}{c^{3}}.
	\end{eqnarray*}
	Using Cauchy Schwartz inequality and Fubini's theorem, we get
	\begin{eqnarray*}
		\Biggl\{ \int_{\mathbb{R}^2 \times \mathbb{R}  \times \mathbb{R}^{+}} |\uptau\textbf{B}_\varphi \psi(c,\alpha,\uptau)|^{p} \frac{dc d\alpha d\uptau}{c^{3}} \Biggr\}^{\frac{1}{p}} \Biggl\{ \int_{\mathbb{R}^2 \times \mathbb{R} \times \mathbb{R}^+}|\gamma \mathfrak{F} \{\textbf{B}_\varphi \psi(c,\alpha,\uptau)\} (\gamma)|^{p}\dfrac{dc d\alpha d\gamma}{c^{3}} \Biggr\}^{\frac{1}{p}}\\\\
		\geq\dfrac{1}{2} \int_{\mathbb{R}^2  \times \mathbb{R}  \times \mathbb{R}^+}  |\textbf{B}_\varphi \psi(c,\alpha,\uptau)|^2 \dfrac{dc d\alpha d\uptau}{c^{3}}.
	\end{eqnarray*}
	Which implies, 
	\begin{equation} \label{8}
		\begin{aligned}
			\Biggl\{ \int_{\mathbb{R}^2 \times \mathbb{R} \times \mathbb{R}^{+}} \bigl|\,\uptau \,\mathbf B_\varphi \psi(c,\alpha,\uptau)\bigr|^{p} \; \frac{dc\;d\alpha\;d\uptau}{c^{3}} \Biggr\}^{\frac{1}{p}}
			\; 
			\Biggl\{ \int_{\mathbb{R}^2 \times \mathbb{R} \times \mathbb{R}^{+}} \bigl|\gamma \,\mathfrak{F} \{ \mathbf B_\varphi \psi(c,\alpha,\uptau)\} (\gamma)\bigr|^{p}\; \frac{dc\;d\alpha\;d\gamma}{c^{3}} \Biggr\}^{\frac{1}{p}} \\
			\ge \; \frac{1}{2}\; \Delta \;\|\psi\|^2.
		\end{aligned}
	\end{equation}
	
	Evaluation of the second integral on the L.H.S of the  above inequality yields
	\begin{align*}
		& \int_{\mathbb{R}^2 \times \mathbb{R} \times \mathbb{R}^+}|\gamma\mathfrak{F} \{\textbf{B}_\varphi \psi(c,\alpha,\uptau)\} (\gamma)|^{p}\dfrac{dc d\alpha d\gamma}{c^{3}}\\\\
		&\qquad\qquad=\int_{\mathbb{R}^2 \times\mathbb{R} \times \mathbb{R}^+ }|\gamma|^{p}|c|^{p}|\hat{\psi}(\gamma)|^{p}  \{ |\hat{\varphi}^{*} (M^{T}_{c,\alpha}\omega)|^{p} +|\check{\hat{\varphi}} (M^{T}_{c,\alpha}\omega)|^{p} \}\frac{dc d\alpha d\uptau}{c^{3}}.
	\end{align*}
	By invoking  Fubini's theorem, we obtain
	\begin{align*}
		&\int_{\mathbb{R}^2 \times \mathbb{R} \times \mathbb{R}^+}|\gamma\mathfrak{F} \{\textbf{B}_\varphi \psi(c,\alpha,\uptau)\} (\gamma)|^{p}\dfrac{dc d\alpha d\gamma}{c^{3}}\\\\
		&=\int_{\mathbb{R}^2}|\gamma|^{p} |\hat{\psi}(\gamma)|^{p}d\gamma \int_{\mathbb{R} \times \mathbb{R}^{+}} |c|^{p} \{ |\hat{\varphi}^{*} (M^{T}_{c,\alpha}\gamma)|^{p} + |\check{\hat{\varphi}} (M^{T}_{c,\alpha}\gamma)|^{p} \}\dfrac{dc d\alpha }{c^{3} }\\\\
		&\leq \int_{\mathbb{R}^2}|\gamma|^{p} |\hat{\psi}(\gamma)|^{p} d\gamma \int_{\mathbb{R} \times \mathbb{R}^{+}} |c|^{2} \biggl\{ \{ |\hat{\varphi}^{*} (M^{T}_{c,\alpha} \gamma)|^{2} + |\check{\hat{\varphi}} (M^{T}_{c,\alpha} \gamma)|^{2} \}\dfrac{dc d\alpha}{c^{3} } \biggr\} ^ {\frac{p}{2}}.
	\end{align*}
	Therefore,
	\begin{align}
		\int_{\mathbb{R}^2 \times \mathbb{R} \times \mathbb{R}^+}|\gamma\mathfrak{F} \{\textbf{B}_\varphi \psi(c,\alpha,\uptau)\} (\gamma)|^{p}\frac{dc d\alpha d\gamma}{c^{3}}\leq \int_{\mathbb{R}^2}|\gamma|^{p} |\hat{\psi}(\gamma)|^{p}d\gamma \times \Delta^{\frac{p}{2}}.
	\end{align}
	Using (8) and (9) and Cauchy Schwartz inequality to obtain,
	\begin{align*}
		&\Biggl\{ \int_{\mathbb{R}^2 \times \mathbb{R} \times \mathbb{R}^{+}} |\uptau\textbf{B}_\varphi \psi(c,\alpha,\uptau)|^{p} \frac{dc d\alpha d\uptau}{c^{3}} \Biggr\}^{\frac{1}{p}} \Biggl\{ \int_{\mathbb{R}^2}|\gamma|^{p} |\hat{\psi}(\gamma)|^{p}d\gamma \times \Delta^{\frac{p}{2}}\Bigg\}^{\frac{1}{p}} \\\\
		& \qquad\qquad  \geq 	\Biggl\{ \int_{\mathbb{R}^2 \times \mathbb{R} \times \mathbb{R}^{+}} |\uptau\textbf{B}_\varphi \psi(c,\alpha,\uptau)|^{p} \frac{dc d\alpha d\uptau}{c^{3}} \Biggr\}^{\frac{1}{p}} \Biggl\{ \int_{\mathbb{R}^2 \times \mathbb{R} \times \mathbb{R}^+}|\gamma\mathfrak{F} \{\textbf{B}_\varphi \psi(c,\alpha,\uptau)\} (\gamma)|^{p}\dfrac{dc d\alpha d\gamma}{c^{3}} \Biggr\}^{\frac{1}{p}} \\\\
		& \qquad\qquad \geq\frac{1}{2}\Delta ||\psi||^2.
	\end{align*}
	Hence,
	\begin{align*}
		\Biggl\{ \int_{\mathbb{R}^2 \times \mathbb{R} \times \mathbb{R}^{+}} |\uptau\textbf{B}_\varphi \psi(c,\alpha,\uptau)|^{p} \frac{dc d\alpha d\uptau}{c^{3}} \Biggr\}^{\frac{1}{p}} \Biggl\{ \int_{\mathbb{R}^2}|\gamma|^{p} |\hat{\psi}(\gamma)|^{p}d\gamma  \Bigg\}^{\frac{1}{p}}\geq \frac{\sqrt{\Delta}}{2}||\psi||^{2}.
	\end{align*}
	
	\parindent=0mm \vspace{0.1in}        
	(ii)	In accordance with the Holder's inequality we may write 
	\begin{align*}
		&\Biggl\{\int_{\mathbb{R}^2 \times \mathbb{R}  \times \mathbb{R}^{+}} |\uptau|^{p}|\textbf{B}_\varphi \psi(c,\alpha,\uptau)|^{2} \frac{dc d\alpha d\uptau}{c^{3}} \Biggr\}^{\frac{2}{p}}  \times 	\Biggl\{\int_{\mathbb{R}^2 \times \mathbb{R} \times \mathbb{R}^{+}} |\textbf{B}_\varphi \psi(c,\alpha,\uptau)|^{2} \frac{dc d\alpha d\uptau}{c^{3}} \Biggr\}^{1-{\frac{2}{p}} }\\\\
		&\quad=	\Biggl\{\int_{\mathbb{R}^2 \times \mathbb{R} \times \mathbb{R}^{+}} \{|\uptau|^{2}|\textbf{B}_\varphi \psi(c,\alpha,\uptau)|^\frac{4}{p}\}^{\frac{p}{2}} \frac{dc d\alpha d\uptau}{c^{3}} \Biggr\}^{\frac{2}{p}}  \times 	\Biggl\{\int_{\mathbb{R}^2 \times \mathbb{R} \times \mathbb{R}^{+}} \{|\textbf{B}_\varphi \psi(c,\alpha,\uptau)|^{2-{\frac{4}{p}}}\}^{\frac{1}{1-\frac{2}{p}}} \frac{dc d\alpha d\uptau}{c^{3}} \Biggr\}^{1-{\frac{2}{p}} }\\\\ 
		&\quad \geq \int_{\mathbb{R}^2 \times \mathbb{R} \times \mathbb{R}^{+}} |\uptau|^{2}|\textbf{B}_\varphi \psi(c,\alpha,\uptau)|^{\frac{4}{p}}  |\textbf{B}_\varphi \psi(c,\alpha,\uptau)|^{2-{\frac{4}{p}}} \frac{dc d\alpha d\uptau}{c^{3}}\\\\
		&\quad=\int_{\mathbb{R}^2 \times \mathbb{R} \times \mathbb{R}^{+}} |\uptau|^{2}|\textbf{B}_\varphi \psi(c,\alpha,\uptau)|^{2} \frac{dc d\alpha d\uptau}{c^{3}}.
	\end{align*}
	
	Therefore, we have
	\begin{align}
		\Biggl\{\int_{\mathbb{R}^2 \times \mathbb{R} \times \mathbb{R}^{+}} |\uptau|^{p}|\textbf{B}_\varphi \psi(c,\alpha,\uptau)|^{2} \frac{dc d\alpha d\uptau}{c^{3}} \Biggr\}^{\frac{1}{p}}  \geq \frac{\Biggl\{ \displaystyle\int_{\mathbb{R}^2 \times \mathbb{R}  \times \mathbb{R}^{+}} |\uptau|^{2}|\textbf{B}_\varphi \psi(c,\alpha,\uptau)|^{2} \frac{dc d\alpha d\uptau}{c^{3}} \Biggr\}^{ \frac{1}{2}}}{ \Biggl\{ \displaystyle\int_{\mathbb{R}^2 \times \mathbb{R}  \times \mathbb{R}^{+}} |\textbf{B}_\varphi \psi(c,\alpha,\uptau)|^{2} \frac{dc d\alpha d\uptau}{c^{3}} \Biggr\}^{\frac{1}{2}-{\frac{1}{p}}}  } .      	\end{align}
	
	In the similar lines as above  and by virtue by Plancherel's formula, we have
	\begin{eqnarray}
		\Biggl\{\int_{\mathbb{R}^2} |\gamma|^{p}|\hat \psi(\gamma)|^{2} d\gamma \Biggr\}^{\frac{1}{p}}  &\geq& \frac{\Biggl\{\displaystyle\int_{\mathbb{R}^2} |\gamma|^{2}|\hat \psi(\gamma)|^{2} d\gamma \Biggr\}^{ \frac{1}{2}}}{ \Biggl\{ \displaystyle\int_{\mathbb{R}^2} |\hat \psi(\gamma)|^{2}d\gamma \Biggr\}^{\frac{1}{2}-{\frac{1}{p}}} } \\&=&   \frac{\Biggl\{\displaystyle\int_{\mathbb{R}^2} |\gamma|^{2}|\hat \psi(\gamma)|^{2} d\gamma \Biggr\}^{ \frac{1}{2}}}{ \Biggl\{ \Delta \times\displaystyle \int_{\mathbb{R}^2} |\hat \psi(\gamma)|^{2}d\gamma \Biggr\}^{\frac{1}{2}-{\frac{1}{p}}} }  \times \Delta^{\frac{1}{2} -{\frac{1}{p}} }\\&=&  \Delta^{\frac{1}{2} -{\frac{1}{p}} } \times\frac{\Biggl\{\displaystyle\int_{\mathbb{R}^2} |\gamma|^{2}|\hat \psi(\gamma)|^{2} d\gamma \Biggr\}^{ \frac{1}{2}}}{ \Biggl\{  \displaystyle\int_{\mathbb{R}^2 \times \mathbb{R}  \times \mathbb{R}^{+}} |\textbf{B}_\varphi \psi(c,\alpha,\uptau)|^{2} \frac{dc d\alpha d\uptau}{c^{3}} \Biggr\}^{\frac{1}{2}-{\frac{1}{p}}}.} 
	\end{eqnarray}
	Multiplying inequalities (10) and (13) to obtain
	{\small
		\begin{align*}
			&\Biggl\{\int_{\mathbb{R}^2 \times \mathbb{R}  \times \mathbb{R}^{+}} |\uptau|^{p}|\textbf{B}_\varphi \psi(c,\alpha,\uptau)|^{2} \frac{dc d\alpha d\uptau}{c^{3}} \Biggr\}^{\frac{1}{p}} \times   \Biggl\{\int_{\mathbb{R}^2} |\gamma|^{p}|\hat \psi(\gamma)|^{2} d\gamma \Biggr\}^{\frac{1}{p}}\\\\
			&\quad\qquad \geq\frac{\Biggl\{\displaystyle\int_{\mathbb{R}^2 \times \mathbb{R} \times \mathbb{R}^{+}} |\uptau|^{2}|\textbf{B}_\varphi \psi(c,\alpha,\uptau)|^{2} \frac{dc d\alpha d\uptau}{c^{3}} \Biggr\}^{ \frac{1}{2}}}{ \Biggl\{ \displaystyle\int_{\mathbb{R}^2 \times \mathbb{R} \times \mathbb{R}^{+}} |\textbf{B}_\varphi \psi(c,\alpha,\uptau)|^{2} \frac{dc d\alpha d\uptau}{c^{3}} \Biggr\}^{\frac{1}{2}-{\frac{1}{p}}} }  \times  \Delta^{\frac{1}{2} -{\frac{1}{p}} }  \times\frac{\Biggl\{\displaystyle\int_{\mathbb{R}^2} |\gamma|^{2}|\hat \psi(\gamma)|^{2} d\gamma \Biggr\}^{ \frac{1}{2}}}{ \Biggl\{ \displaystyle \int_{\mathbb{R}^2 \times \mathbb{R} \times \mathbb{R}^{+}} |\textbf{B}_\varphi \psi(c,\alpha,\uptau)|^{2} \frac{dc d\alpha d\uptau}{c^{3}} \Biggr\}^{\frac{1}{2}-{\frac{1}{p}}} } \\\\
			&\qquad\quad = \Delta^{\frac{1}{2} -{\frac{1}{p}} } \times  \frac{\Biggl\{\displaystyle\int_{\mathbb{R}^2 \times \mathbb{R} \times \mathbb{R}^{+}} |\uptau|^{2}|\textbf{B}_\varphi \psi(c,\alpha,\uptau)|^{2} \frac{dc d\alpha d\uptau}{c^{3}} \Biggr\}^{ \frac{1}{2}} \times \Biggl\{\displaystyle\int_{\mathbb{R}^2} |\gamma|^{2}|\hat \psi(\gamma)|^{2} d\gamma \Biggr\}^{\frac{1}{2}}}{ \Biggl\{\displaystyle \int_{\mathbb{R}^2 \times \mathbb{R} \times \mathbb{R}^{+}} |\textbf{B}_\varphi \psi(c,\alpha,\uptau)|^{2} \frac{dc d\alpha d\uptau}{c^{3}} \Biggr\}^{1-{\frac{2}{p}}} } \\\\
			&\qquad\quad \geq \frac{1}{2} \frac{ \Delta^{\frac{1}{2} -{\frac{1}{p}} } \times \sqrt{\Delta} ||\psi||^2}{ [\Delta ||\psi||^2 ]^{1-{\frac{2}{p}}}}\\\\
			& \qquad\quad=\frac{1}{2} \Delta^{\frac{1}{p}} ||\psi||^{\frac{4}{p}}.~~~\square
	\end{align*}}

	\parindent=0mm \vspace{0.1in}
	{\bf{Remark 3.2.}} If we put $p=2$ into inequalities 6 and 7 theorem (4.2)
	transforms into classical Heisenberg's uncertainty principle for the boostlet transform, given by
	\begin{align*}
		\Biggl\{ \int_{\mathbb{R}^2 \times \mathbb{R}  \times \mathbb{R}^{+}} |\uptau|^2 |\textbf{B}_\varphi \psi(c,\alpha,\uptau)|^2 \frac{dc d\alpha d\uptau}{c^{3}} \Biggr\}^{\frac{1}{2}}\Biggl\{ \int_{\mathbb{R}^2}|\gamma|^2 |\hat{\psi}(\gamma)|^2 d\gamma \Biggr\}^\frac{1}{2} \geq \frac{1}{2} \sqrt{\Delta} ||\psi||^2.
	\end{align*}   
	
	\parindent=8mm \vspace{0.1in}
	Before establishing the logarithmic uncertainty inequality, we need to define the Schwartz space in $ L^2(\mathbb{R}^2) $ denoted by $ \mathbb{S}(\mathbb{R}^2) $, as the domain of functions for which the inequality holds. The elements of this space are infinitely differentiable functions that, along with all their derivatives, decay faster than any inverse power of $|x|$ as $|x| \rightarrow \infty $ and can be defined as:
	
	\begin{equation*}
		\mathbb{S}(\mathbb{R}^2) = \Biggl\{ f \in \mathbb{C}^{\infty} (\mathbb{R}^2) : \sup_{x \in \mathbb{R}^2} \big| x^{\alpha} {\partial_x}^{\beta} f(x) \big| \Biggr\}
	\end{equation*} 
	where $ \mathbb{C}^{\infty}(\mathbb{R}^2) $ is the class of infinitely differentiable functions, with $ \alpha, \beta $ denoting the multi-indices and $  \partial_x $ is  the standard partial differential operator with respect to variable x. We can now proceed to derive the logarithmic uncertainty principle for the continuous boostlet transform.
	
	\parindent=0mm \vspace{0.1in}   		
	{\bf{Theorem 3.3.}}
	{\textbf{(Logarithmic Inequality)}}
	Let  $ \varphi \in L^2(\mathbb{R}^2) $  be an admissible boostlet and $  \textbf{B}_\varphi \psi(c,\alpha,\uptau) $ denotes the boostlet transform of any non-trivial function $ \psi \in \mathbb{S}(\mathbb{R}^2)\subseteq {L^2(\mathbb{R}^2)}$,then the inequality given below is true:
	\begin{align*}
		\int_{\mathbb{S}} \ln|\uptau| |\textbf{B}_\varphi \psi(c,\alpha,\uptau)|^{2} \dfrac{dc d\alpha d\uptau}{c^{3}} + \Delta \int_{\mathbb{R}^2} \ln|\gamma| |\hat \psi(\gamma)|^2 d\gamma  \geq \Delta ||\psi||^2 \Bigg[ \frac{ \Gamma'\big( \frac{1}{2} \big)}{  \Gamma \big( \frac{1}{2} \big)} -\ln \pi \Bigg].
	\end{align*}
	
	\parindent=0mm \vspace{.1in}
	\textbf{Proof}. 
	For a non-trivial function $\psi \in \mathbb{S}(\mathbb{R}^2)$,the classical  Logarithmic inequality in time and frequency domain is given by~\cite{bw}
	\begin{align*}
		\int_{\mathbb{R}^2} \ln|t| | \psi(t)|^{2} dt +  \int_{\mathbb{R}^2} \ln|\gamma| |\hat \psi(\gamma)|^2 d\gamma  \geq \Bigg[ \frac{ \Gamma'\big( \frac{1}{2} \big)}{  \Gamma \big( \frac{1}{2} \big)} -\ln \pi \Bigg] \int_{\mathbb{R}^2}|\psi(t)|^2 dt
	\end{align*}
	Replacing $ \psi $ by $ \textbf{B}_\varphi \psi(c,\alpha,\uptau) $ so that 
	\begin{align*}
		\int_{\mathbb{R}^2} \ln|\uptau| |\textbf{B}_\varphi \psi(c,\alpha,\uptau) |^{2} d\uptau  +  \int_{\mathbb{R}^2} \ln|\gamma| |\mathfrak{F} \{ \textbf{B}_\varphi \psi(c,\alpha,\uptau) \}(\gamma)|^2 d\gamma\\ \geq \Bigg[ \frac{ \Gamma'\big( \frac{1}{2} \big)}{ \Gamma \big( \frac{1}{2} \big)} -\ln\pi \Bigg] \int_{\mathbb{R}^2}|\textbf{B}_\varphi \psi(c,\alpha,\uptau) |^2 d\uptau
	\end{align*}
	Integrating w.r.t the measure $ \dfrac{dc d\alpha}{c^{3}}  $ and using
	Plancherel's formula ,we have
	\begin{align*}
		\int_{\mathbb{R} \times \mathbb{R}^{+}} \Biggl\{ \int_{\mathbb{R}^2} \ln|\uptau| |\textbf{B}_\varphi \psi(c,\alpha,\uptau) |^{2} d\uptau  +  \int_{\mathbb{R}^2} \ln|\gamma| |\mathfrak{F} \{ \textbf{B}_\varphi \psi(c,\alpha,\uptau) \}(\gamma)|^2 d\gamma  \Biggr\} \frac{dc d\alpha}{c^{3}} \\ \geq \Bigg[ \frac{ \Gamma'\big( \frac{1}{2} \big)}{ \Gamma \big( \frac{1}{2} \big)} -\ln\pi \Bigg] \int_{\mathbb{R}^2  \times \mathbb{R} \times \mathbb{R}^+} |\textbf{B}_\varphi \psi(c,\alpha,\uptau) |^2 \frac{dc d\uptau d\alpha}{c^3}   	
	\end{align*}
	
	\begin{align}
		\int_{\mathbb{R}^2  \times \mathbb{R} \times \mathbb{R}^+}  \ln|\uptau| |\textbf{B}_\varphi \psi(c,\alpha,\uptau) |^{2} \frac{dc d\alpha d\uptau}{c^{3}} + 	\int_{\mathbb{R}^2  \times \mathbb{R} \times \mathbb{R}^+} \ln|\gamma| |\mathfrak{F} \{ \textbf{B}_\varphi \psi(c,\alpha,\uptau) \}(\gamma)|^2 \frac{dc d\alpha d\gamma}{c^{3}} \\ \geq \Bigg[ \frac{ \Gamma'\big( \frac{1}{2} \big)}{ \Gamma \big( \frac{1}{2} \big)} -\ln\pi \Bigg] \Delta ||\psi||^2
	\end{align} 
	 Evaluating  second integral on the L.H.S of the above inequality yields  
	\begin{align*}
		&\int_{\mathbb{R}^2  \times \mathbb{R} \times \mathbb{R}^+} \ln|\gamma| |\mathfrak{F} \{ \textbf{B}_\varphi \psi(c,\alpha,\uptau) \}(\gamma)|^2 \frac{dc d\alpha d\gamma}{c^{3}} \\	&\quad \quad\quad\quad \quad\quad= \int_{\mathbb{R}^2  \times \mathbb{R} \times \mathbb{R}^+}\ln|\gamma| |c|^2 |\hat{\psi}(\gamma)|^2 \Big[ |{\hat{\varphi}}^{*}(M^{T}_{c,\alpha}\gamma) |^2 + |\check{\hat{\varphi}}(M^{T}_{c,\alpha}\gamma) |^2    \Big] \frac{dc d\alpha d\gamma}{c^{3}} \\
		\\	&\quad \quad\quad\quad \quad\quad=\int_{\mathbb{R}^2}\ln|\gamma| |\hat{f}(\gamma)|^2 d\gamma \int_{\mathbb{R} \times \mathbb{R}^{+}} \Big[ |{\hat{\varphi}}^{*}(M^{T}_{c,\alpha}\gamma) |^2 + |\check{\hat{\varphi}}(M^{T}_{c,\alpha}\gamma) |^2 \Big] \frac{dc d\alpha}{c}\\	&\quad \quad\quad\quad \quad\quad=
		\int_{\mathbb{R}^2} \ln|\gamma| |\hat{\psi}(\gamma)|^2 d\gamma \big[ \Delta _\varphi^{*}(\gamma) +\Delta _\varphi (\gamma) \big] 
		\\	&\quad \quad\quad\quad \quad\quad=\Delta	\int_{\mathbb{R}^2} \ln|\gamma| |\hat{\psi}(\gamma)|^2 d\gamma
	\end{align*}
	Using this in inequality (14) we get   
	\begin{align*}
		\int_{\mathbb{S}} \ln|\uptau| |\textbf{B}_\varphi \psi(c,\alpha,\uptau)|^{2} \dfrac{dc d\alpha d\uptau}{c^{3}} + \Delta \int_{\mathbb{R}^2} \ln|\gamma| |\hat \psi (\gamma)|^2 d\gamma \ge  \Delta ||\psi||^2 \Bigg[ \frac{ \Gamma'\big( \frac{1}{2} \big)}{  \Gamma \big( \frac{1}{2} \big)} -\ln \pi \Bigg].~~\square
	\end{align*}

	\parindent=0mm \vspace{0.1in}
	{\bf{Theorem 3.4.}}	
	{\textbf{(Pitt's Inequality)}}
	Let $ f \in \mathbb{S}(\mathbb{R}^2) $ be such that $  \textbf{B}_\varphi f(c,\alpha,\uptau) \in \mathbb{S}(\mathbb{R}^2), $ where $ \varphi $ is an admissible boostlet. Then the following inequality holds:
	\begin{align*}
		\Delta \int_{\mathbb{R}^2} |\omega|^{-\lambda} |\hat{f}(\omega)|^2 d\omega \leq C_\lambda \int_{\mathbb{S}} |\uptau|^{\lambda} | \textbf{B}_\varphi f(c,\alpha,\uptau)|^2 \frac{dc d\alpha d\uptau}{c^3}, 
	\end{align*}
	Where,\begin{equation*}
		C_{\lambda} = \pi^{\lambda}\Bigg[ \Gamma\big( \frac{2-\lambda}{4} \big)  /  \Gamma \big( \frac{2+\lambda}{4} \big) \Bigg]^{2}.
	\end{equation*}
	\parindent=0mm \vspace{.1in}
	\textbf{Proof.} For any $ f \in  \mathbb{S}(\mathbb{R}^2) \subseteq L^2(\mathbb{R}^2) $ the classical Pitt's inequality is given as~\cite{bw}
	\begin{align*}
		\int_{\mathbb{R}^2} |\omega|^{-\lambda} |\hat{f}(\omega)|^2 d\omega \leq C_\lambda \int_{\mathbb{R}^2} |x|^{\lambda} | f(x)|^2 dx ; 0\leq\lambda < 2,
	\end{align*}
	where,\begin{equation*}
		C_{\lambda} = \pi^{\lambda}\Bigg[ \Gamma\big( \frac{2-\lambda}{4} \big)  /  \Gamma \big( \frac{2+\lambda}{4} \big) \Bigg]^{2}.
	\end{equation*}
	Replacing $f$ by $ \textbf{B}_\varphi f(c,\alpha,\uptau) $, so that
	\begin{align*}
		\int_{\mathbb{R}^2} |\omega|^{-\lambda} |\mathcal{F} \{ \textbf{B}_\varphi f(c,\alpha,\uptau) \} (\omega)|^2 d\omega \leq C_\lambda \int_{\mathbb{R}^2} |\uptau|^{\lambda} |\textbf{B}_\varphi f(c,\alpha,\uptau) |^2 d\uptau.  
	\end{align*}
	Integrating w.r.t the measure $\dfrac{dc d\alpha}{c^{3}}$ and using Fubini's theorem, we have
	\begin{align}
		&\int_{\mathbb{R}^2  \times \mathbb{R} \times \mathbb{R}^+}|\omega|^{-\lambda} |\mathcal{F} \{ \textbf{B}_\varphi f(c,\alpha,\uptau) \} (\omega)|^2 \dfrac{dc d\alpha d\omega}{c^3}\\
		&\qquad\qquad\qquad\qquad \leq C_\lambda \int_{\mathbb{R}^2 \times \mathbb{R} \times \mathbb{R}^{+}} |\uptau|^{\lambda} |\textbf{B}_\varphi f(c,\alpha,\uptau) |^2 \dfrac{dc d\alpha d\uptau}{c^3}
	\end{align}
	The integral on L.H.S of above inequality can be evaluated as 
	\begin{align*}
		&\int_{\mathbb{R}^2  \times \mathbb{R} \times \mathbb{R}^+}|\omega|^{-\lambda} |\mathcal{F} \{ \textbf{B}_\varphi f(c,\alpha,\uptau) \} (\omega)|^2 \dfrac{dc d\alpha d\omega}{c^3}\\\\
		& \qquad\qquad= \int_{\mathbb{R}^2  \times \mathbb{R} \times \mathbb{R}^+} |\omega|^{-\lambda}|c|^{2}|\hat{f}(\omega)|^{2}\Big[ |{\hat{\varphi}}^{*}(M^{T}_{c,\alpha}\omega) |^2 + |\check{\hat{\varphi}}(M^{T}_{c,\alpha}\omega) |^2    \Big] \dfrac{dc d\alpha d\omega}{c^{3}}\\\\
		&\qquad \qquad = \int_{\mathbb{R}^2}|\omega|^{-\lambda} |\hat{f}(\omega)|^2 d\omega \int_{\mathbb{R} \times \mathbb{R}^{+}} \Big[ |{\hat{\varphi}}^{*}(M^{T}_{c,\alpha}\omega) |^2 + |\check{\hat{\varphi}}(M^{T}_{c,\alpha}\omega) |^2 \Big] \dfrac{dc d\alpha}{c}\\\\
		&\qquad \qquad =	\int_{\mathbb{R}^2} |\omega|^{-\lambda} |\hat{f}(\omega)|^2 d\omega \big[ \Delta _\varphi^{*}(\omega) +\Delta _\varphi (\omega) \big] \\\\
		&\qquad \qquad=\Delta	\int_{\mathbb{R}^2} |\omega|^{-\lambda} |\hat{f}(\omega)|^2 d\omega
	\end{align*}
	Using this in (16), we get 
	\begin{align*}
		\Delta \int_{\mathbb{R}^2} |\omega|^{-\lambda} |\hat{f}(\omega)|^2 d\omega \leq C_\lambda \int_{\mathbb{S}} |\uptau|^{\lambda} | \textbf{B}_\varphi f(c,\alpha,\uptau)|^2 \dfrac{dc d\alpha d\uptau}{c^3}. 
	\end{align*}
	Thus the proof is completed.~~~$ \square$
	\parindent=0mm \vspace{0.1in}
	
	{\bf{Theorem 3.5.}}	
	{(\textbf{Nazarov's Inequality})}
	Let $ \varphi \in L^2(\mathbb{R}^2) $ be an admissible boostlet and $ A_1,A_2 $ be two measurable finite subsets of $\mathbb{R}^2 $.Then for every function $f \in  L^2(\mathbb{R}^2) $ such that $\textbf{B}_\varphi f(c,\alpha,\uptau) \in  L^2(\mathbb{R}^2), $ we have
	\begin{align}
		\Delta ||f||^2 \leq M e^{M(A_1,A_2)}\Biggl\{ \Delta||f||^2 -\int_{\mathbb{R} \times{\mathbb R}^+ \times{A_1}} |\textbf{B}_\varphi f(c,\alpha,\uptau)|^2\, \dfrac{dc d\alpha d\uptau}{c^{3}}\\ \qquad\qquad\qquad+\Delta \int_{\mathbb{R}^2} |\hat{f}(\omega)|^2 d\omega -\Delta \int_{A_2}|\hat{f}(\omega)|^2d\omega \Biggr\},
	\end{align}
	where,
	\begin{align*}
		M e^{M(A_1,A_2)}=M \min \big( |A_1| |A_2|,|A_1|^{\frac{1}{n}} w(A_2) ,w(A_ 1)|A_2| \big);
	\end{align*} 
	\indent   $ w(A_i) ,i=1,2$  is the mean width of $ A_i $ and $ |A_i| , i=1,2 $	is the Lebesgue measure of $ A_i $.
	\parindent=0mm \vspace{0.1in}
	
	\textbf{Proof}. 
	From Nazarov's uncertainty principle for the FT we get
	\begin{align}
		\int_{\mathbb{R}^2}|f(t) |^2 dt \leq M e^{M(A_1,A_2)}\Biggl\{   \int_{\mathbb{R}^2 \setminus A_1}|f(t)|^2 dt + \int_{\mathbb{R}^2 \setminus A_2} |\hat{f}(\omega)^2 d\omega \Biggr\}
	\end{align}
	Where 	\begin{align*}
		M e^{M(A_1,A_2)}=M min \big( |A_1| |A_2|,|A_1|^{\frac{1}{n}} w(A_2) ,w(A_ 1)|A_2| \big),
	\end{align*} 
	\indent   $ w(A_i) ,i=1,2$  is the mean width of $ A_i $ and $ |A_i| , i=1,2 $	is the Lebesgue measure of $ A_i .$
	Replacing $ f $ by  $\textbf{B}_\varphi f(c,\alpha,\uptau) $ we see that 
	\begin{align*}
		&\int_{\mathbb{R}^2}|\textbf{B}_\varphi f(c,\alpha,\uptau) |^2 d\uptau  \leq M e^{M(A_1,A_2)}\Biggl\{   \int_{\mathbb{R}^2 \setminus A_1}| \textbf{B}_\varphi f(c,\alpha,\uptau)|^2 d\uptau \\\
		&\qquad\qquad\qquad\qquad\qquad\qquad\qquad\qquad\qquad\qquad+ \int_{\mathbb{R}^2 \setminus A_2} |\mathcal{F} \{ \textbf{B}_\varphi f(c,\alpha,\uptau)\}(\omega)|^2 d\omega \Biggr\}.
	\end{align*} 
	Integrating w.r.t the measure $ \dfrac{dc d\alpha}{c^{3}}  $ and using
	Plancherel's formula ,we find that 
	\begin{align*}
		\int_{\mathbb{R} \times{\mathbb R}^{+}}	\int_{\mathbb{R}^2}|\textbf{B}_\varphi f(c,\alpha,\uptau) |^2   \dfrac{dc d\uptau d\alpha}{c^{3}}& \leq M e^{M(A_1,A_2)}\Biggl\{  \int_{\mathbb{R} \times{\mathbb R}^{+}}   \Bigg( \int_{\mathbb{R}^2 \setminus A_1}| \textbf{B}_\varphi f(c,\alpha,\uptau)|^2 d\uptau \\\\
		&\qquad  \quad+ \int_{\mathbb{R}^2 \setminus A_2} |\mathcal{F} \{ \textbf{B}_\varphi f(c,\alpha,\uptau)\}(\omega)|^2 d\omega \Bigg)  \dfrac{dc d\alpha}{c^{3}}   \Biggr\},
	\end{align*}
	which implies,
	\begin{align*}
		&\int_{\mathbb{R}^2  \times {\mathbb R} \times{\mathbb R}^+}|\textbf{B}_\varphi f(c,\alpha,\uptau) |^2 \dfrac{dc d\alpha d\uptau}{c^3}\\\\
		&\qquad \leq  M e^{M(A_1,A_2)} \Biggl\{ \int_{\mathbb{R}^2  \times {\mathbb R} \times{\mathbb R}^+}|\textbf{B}_\varphi f(c,\alpha,\uptau) |^2\,  \dfrac{dc d\alpha d\uptau}{c^3} -  \int_{ {\mathbb R} \times{\mathbb R}^+ \times{A_1}} |\textbf{B}_\varphi f(c,\alpha,\uptau) |^2\,  \dfrac{dc d\alpha d\uptau}{c^3}\\\\
		&\qquad\qquad +	\int_{\mathbb{R}^2  \times {\mathbb R} \times{\mathbb R}^+} |\mathcal{F} \{ \textbf{B}_\varphi f(c,\alpha,\uptau)\}(\omega)|^2  \frac{dc d\alpha d\omega}{c^3} -  \int_{\mathbb{R} \times{\mathbb R}^+ \times{A_2}} |\mathcal{F} \{ \textbf{B}_\varphi f(c,\alpha,\uptau)\}(\omega)|^2\,  \dfrac{dc d\alpha d\omega}{c^3}  \Biggr\}.
	\end{align*} 
	Therefore, we have 
	\begin{align*}
		&  \Delta ||f||^2 \leq  M e^{M(A_1,A_2)}
		\Biggl\{  \Delta ||f||^2  -  \int_{ {\mathbb R} \times{ \mathbb R}^+ \times{A_1}} |\textbf{B}_\varphi f(c,\alpha,\uptau) |^2 \dfrac{dc d\alpha d\uptau}{c^3}\\\\ 
		&\qquad\qquad\qquad\qquad\qquad+  \int_{\mathbb{R}^2  \times {\mathbb R} \times{ \mathbb R}^+} |c|^2 |\hat{f}(\omega)|^2 \Big[ |{\hat{\varphi}}^{*}(M^{T}_{c,\alpha}\omega) |^2 + |\check{\hat{\varphi}}(M^{T}_{c,\alpha}\omega) |^2    \Big] \dfrac{dc d\alpha d\uptau}{c^{3}} \\\\
		&\qquad\qquad\qquad\qquad\qquad \qquad\qquad\qquad\qquad\qquad\qquad-  \int_{\mathbb{R} \times{\mathbb R}^+ \times{A_2}} |\mathcal{F} \{ \textbf{B}_\varphi f(c,\alpha,\uptau)\}(\omega)|^2  \dfrac{dc d\alpha d\omega}{c^3}  \Biggr\} \\\\
		&\qquad \quad\leq  M e^{M(A_1,A_2)}
		\Biggl\{  \Delta ||f||^2  -  \int_{\mathbb{R} \times{\mathbb R}^+ \times{A_1}} |\textbf{B}_\varphi f(c,\alpha,\uptau) |^2 \frac{dc d\alpha d\uptau}{c^3} +  \Delta \int_{\mathbb{R}^2} |\hat{f}(\omega)|^2 d\omega \\\\
		&\qquad\qquad\qquad\qquad\qquad\qquad\qquad\quad -  \int_{\mathbb{R} \times{\mathbb R}^+ \times{A_2}}   |c|^2 |\hat{f}(\omega)|^2 \Big[ |{\hat{\varphi}}^{*}(M^{T}_{c,\alpha}\omega) |^2 + |\check{\hat{\varphi}}(M^{T}_{c,\alpha}\omega) |^2 \dfrac{dc d\alpha d\omega}{c^3}  \Biggr\} \\\\
		&\qquad\quad  \leq  M e^{M(A_1,A_2)}
		\Biggl\{  \Delta ||f||^2  -  \int_{ {\mathbb R} \times{\mathbb R}^+ \times{A_1}} |\textbf{B}_\varphi f(c,\alpha,\uptau) |^2 \dfrac{dc d\alpha d\uptau}{c^3} +  \Delta \int_{\mathbb{R}^2} |\hat{f}(\omega)|^2 d\omega \\\\
		&\qquad\qquad\qquad\qquad\qquad\qquad\qquad\qquad - \int_{\mathbb{R} \times{\mathbb R}^{+}} \bigg( |{\hat{\varphi}}^{*}(M^{T}_{c,\alpha}\omega) |^2 + |\check{\hat{\varphi}}(M^{T}_{c,\alpha}\omega) |^2\bigg) \dfrac{dc d\alpha}{c} \int_{A_2}|\hat{f}(\omega)|^2 d\omega \Biggr\}. 
	\end{align*}
	Thus we have 
	{\small
		\begin{align*}
			\Delta ||f||^2 \leq  M e^{M(A_1,A_2)}
			\Biggl\{  \Delta ||f||^2  -  \int_{ {\mathbb R} \times{\mathbb R}^+ \times{A_1}} |\textbf{B}_\varphi f(c,\alpha,\uptau) |^2 \dfrac{dc d\alpha d\uptau}{c^3} +  \Delta \int_{\mathbb{R}^2} |\hat{f}(\omega)|^2 d\omega  - \Delta \int_{A_2}|\hat{f}(\omega)|^2 d\omega  \Biggr\} .
		\end{align*}		
		
		This completes the proof.
		
		\parindent=0mm \vspace{0.1in}  
		{\bf{4. Example }}
		
		\parindent=0mm \vspace{0.1in}
		
		This section presents a detailed computation of the boostlet transform applied to the exponential function $ f(\mu) = e^{||\mu||^2} $ in 2D Minkowski space with metric $ \eta = \mathrm{diag}(1,-1) $. Using an admissible window function tailored for the Lorentzian metric, the transform incorporates dilation, hyperbolic rotation, and spacetime translation via the Lorentz boost matrix $ M_{c,\alpha} $. The closed-form expressions of the transform components reveal how boostlets localize and represent Lorentz-invariant structures in spacetime. Accompanied by an interactive surface plot of the boostlet transform magnitude, this illustrates the effects of Lorentz boosts, dilation factors, and spacetime shifts on the transform’s behavior, providing insight into its analytic and geometric properties. These results advance the theoretical understanding and potential applications of boostlet transforms in spacetime harmonic analysis.

		{\bf{Example 4.1}} Consider a function $ f(\mu) = e^{||\mu||^2} $ and the window function $\varphi(\mu)= \exp\biggl\{ -\frac{1}{2} (s^2-(1- \iota \epsilon))t^2  \biggr\}    $, where $ \epsilon >0$,  defined in 2D Minkowski space with metric $\eta= \mathrm{diag}(1,-1) $.    Then the boostlet transform of the function $ f(\mu) = e^{||\mu||^2} $, is calculated as follows :
		\begin{equation*}
			\textbf{B}_\varphi f(c,\alpha,\uptau)= \left( \langle f,\varphi_{c,\alpha,\uptau}\rangle,\langle f,\varphi^{*}_{c,\alpha,\uptau}\rangle\right), 
		\end{equation*}
		We have
		{\small 
			\begin{eqnarray*}
				\langle f,\varphi_{c,\alpha,\uptau}\rangle&=&\int_{\mathbb{R}^2} f(\mu) \varphi^{*}_{c,\alpha,\uptau} d\mu\\&=&c^{-1} \int_{\mathbb{R}^2} f(\mu) \varphi^{*}(M_{c,\alpha}^{-1}(\mu-\uptau))d\mu.
			\end{eqnarray*}
				Put $ M_{c,\alpha}^{-1}(\mu-\uptau) =\nu	$, so that $ \mu= M_{c,\alpha} \nu +\uptau $ and $ d\mu= | M_{c,\alpha}| d\nu $, which implies $ d\mu= c^2 d\nu $.

		Therefore, 
		\begin{eqnarray*}
			\langle f,\varphi_{c,\alpha,\uptau}\rangle&=& c^{-1} \int_{\mathbb{R}^2} f( M_{c,\alpha} \nu +\uptau)\varphi^{*}(\nu)c^2 d\nu\\ 
			 & =& c  \int_{\mathbb{R}^2} \exp^{||M_{c,\alpha} \nu +\uptau||^2 } \exp\biggl\{ -\frac{1}{2} (\nu_{s}^2-(1+ \iota \epsilon))\nu_{t}^2  \biggr\} d\nu.
		\end{eqnarray*}
		Since,
		\begin{eqnarray*}
			||M_{c,\alpha} \nu +\uptau||^2&=&(M_{c,\alpha} \nu +\uptau)^T\eta (M_{c,\alpha} \nu +\uptau)\\&=&\nu^{T} M_{c,\alpha}^{T} M_{c,\alpha}  \nu +\nu^{T} M_{c,\alpha}^{T} \eta \uptau + \uptau^{T} \eta M_{c,\alpha} \nu +\uptau^{T} \eta \uptau \\&=&\nu^{T}Q\nu +L^{T} \nu + C.
		\end{eqnarray*}
		where,
		$ Q= M_{c,\alpha}^{T}\eta M_{c,\alpha}=c^2 B_{\alpha}^{T}\eta B_{\alpha} =c^2 \eta $, since $B_{\alpha}^{T}\eta B_{\alpha}=\eta $,
		$ L=2 M_{c,\alpha}^{T} \eta  \uptau =2cB_{\alpha} \eta \uptau $,
		$C = \uptau^{T} \eta \uptau= \uptau_{s}^2 -\uptau_{t}^2$.\\
		Therefore 
		\begin{eqnarray*}
			\langle f,\varphi_{c,\alpha,\uptau}\rangle&=&c\int_{\mathbb{R}^2} \exp \biggl\{ \nu^{T}Q\nu +L^{T} \nu + \uptau^{T} \eta \uptau +  -\frac{1}{2} (\nu_{s}^2-(1+\iota \epsilon)\nu_{t}^2)       \biggr\} d\nu\\&=&c\int_{\mathbb{R}^2} \exp \biggl\{ -\frac{1}{2}\nu^{T}Q_{1}\nu+ +L^{T} \nu + C  \biggr\} d\nu
		\end{eqnarray*}
		where 
		\begin{eqnarray*}
		Q_{1}&=& -2Q + \begin{pmatrix}1&0\\0& -(1+\iota \epsilon)\end{pmatrix}\\ &=&\begin{pmatrix}-2c^2&0\\0& 2c^2\end{pmatrix}+\begin{pmatrix}1&0\\0& -(1+\iota \epsilon)\end{pmatrix}\\&=& \begin{pmatrix}1-2c^2&0\\0&2c^2 -(1+\iota \epsilon)\end{pmatrix}.
		\end{eqnarray*} 
		
	Therefore, 
	\begin{eqnarray*}
		\langle f,\varphi_{c,\alpha,\uptau}\rangle&=&c \exp \biggl\{ C+\frac{1}{2} L^{T}Q_1^{-1} L \biggr\} \frac{2 \pi}{\sqrt{detQ_{1} }}\\&=&ce^{\uptau_{s}^2-\uptau_{t}^2}\frac{2 \pi}{\sqrt{(1-2c^2)(2c^2-1-\iota \epsilon) }}\exp \biggl\{ 2c^2 \biggl(    \frac{ L_{s}^2}{1-2c^2}+\frac{ L_{t}^2}{2c^2-1-\iota \epsilon}\biggr) \biggr\} 
	\end{eqnarray*}
	where, $ L_{s}= \cosh \alpha  \uptau_{s}+\sinh \alpha \uptau_{t},$ and 
	$ L_{t}= -\sinh \alpha  \uptau_{s}-\cosh \alpha \uptau_{t}$.
	
	Similarly,
	\begin{equation*}
		\langle f,\varphi^{*}_{c,\alpha,\uptau}\rangle=c e^{-\uptau_{s}^2+\uptau_{t}^2}\frac{2 \pi}{\sqrt{(1-2c^2)(2 c^2-1-\iota \epsilon) }}\exp \biggl\{ 2 c^2 \biggl( \frac{ L_{s1}^2 }{1-2 c^2}+\frac{ L_{t1}^2}{2 c^2-1-\iota \epsilon} \biggr) \biggr\} ,
	\end{equation*}
	where, $ L_{s1}= -\cosh \alpha  \uptau_{s}+\sinh \alpha \uptau_{t},$  and 
	$ L_{t1}= \sinh \alpha  \uptau_{s}+\cosh \alpha \uptau_{t}$.\\
	
Thus we have
		\begin{align*}
			\mathbf B_{\varphi}f(c,\alpha,\uptau)
			&=
			\Biggl(
			\; c\,e^{\uptau_{s}^2 - \uptau_{t}^2}
			\;\frac{2\pi}
			{\sqrt{\,(1 - 2c^2)\,(2c^2 - 1 - \mathrm{i}\,\epsilon)\,}}
			\;\exp\!\Bigl\{
			2c^2 \Bigl(
			\frac{L_{s}^2}{1 - 2c^2}
			+ \frac{L_{t}^2}{2c^2 - 1 - \mathrm{i}\,\epsilon}
			\Bigr)
			\Bigr\}, \\
			& \qquad
			c\,e^{-\uptau_{s}^2 + \uptau_{t}^2}
			\;\frac{2\pi}
			{\sqrt{\,(1 - 2c^2)\,(2c^2 - 1 - \mathrm{i}\,\epsilon)\,}}
			\;\exp\!\Bigl\{
			2c^2 \Bigl(
			\frac{L_{s1}^2}{1 - 2c^2}
			+ \frac{L_{t1}^2}{2c^2 - 1 - \mathrm{i}\,\epsilon}
			\Bigr)
			\Bigr\}
			\Biggr).
		\end{align*}
	
	\begin{figure}[H]
		\centering
		\includegraphics[width=1.2\textwidth]{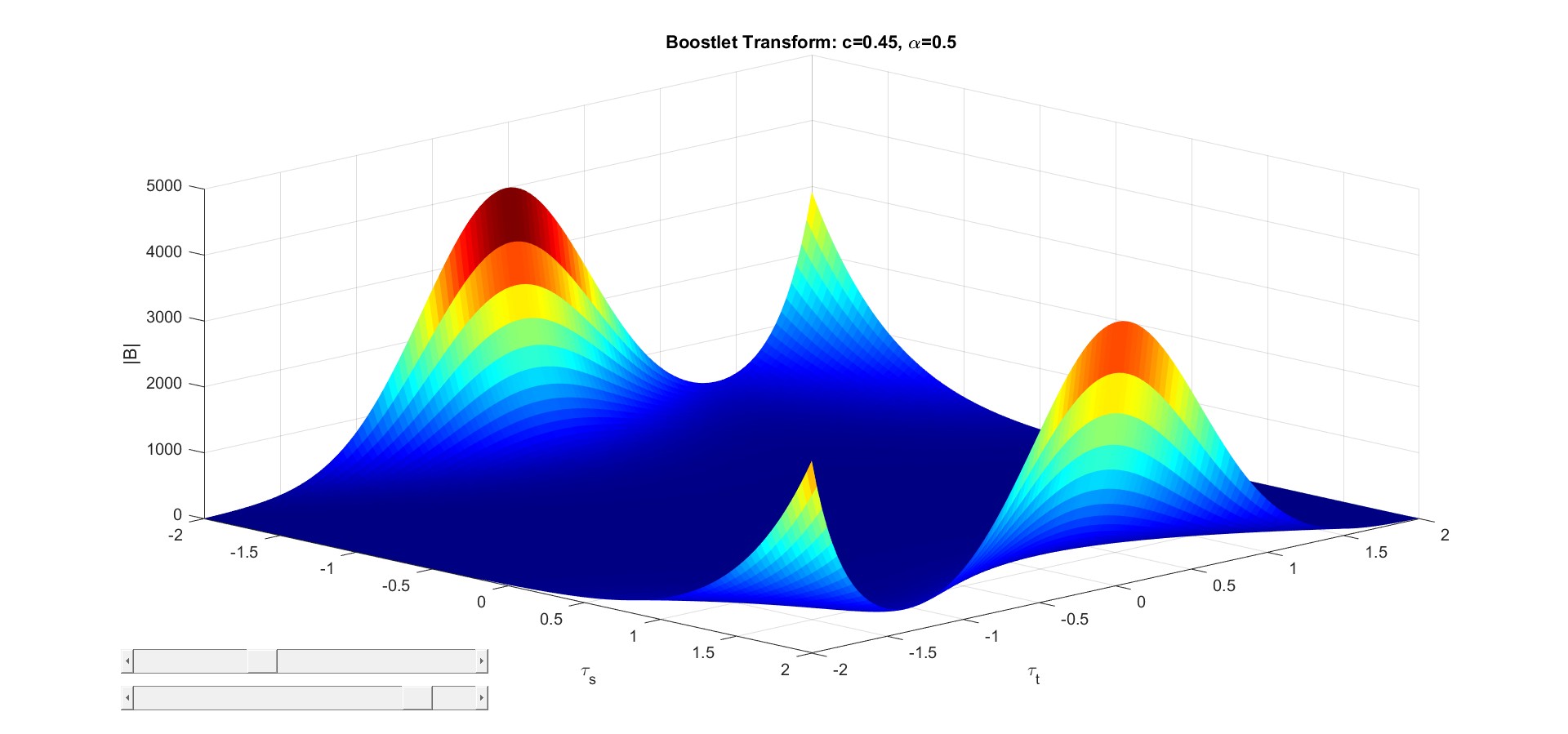}
		\caption{Interactive surface plot of the Boostlet transform magnitude 
			\( |B_{\varphi}f(c,\alpha,\tau)| \) in 2D Minkowski space, showing how 
			the transform varies under different Lorentz boost parameters (\(\alpha\)), 
			dilation factors (\(c\)), and spacetime translations 
			(\(\tau = (\tau_{s}, \tau_{t})\)). The slider controls enable exploration 
			of boost-induced hyperbolic deformation effects and their impact on the 
			localization of boostlet energy in spacetime.}
		\label{fig:boostlet_transform_interactive}
	\end{figure}

		\parindent=0mm \vspace{0.1in}      
		{\bf{5. Potential Applications}}
		
		\parindent=0mm \vspace{0.1in}      
		The Boostlet Transform offers a highly adaptable framework for time-frequency analysis, making it particularly well-suited for applications where signals exhibit time-varying, transient, or non-stationary behaviors. By providing flexible time-frequency resolution, the Boostlet Transform is capable of optimizing the analysis of complex signals with varying spectral content. Below are several key potential applications where the Boostlet Transform can be effectively utilized:
		
		\parindent=0mm \vspace{0.1in}
		{\it 5.1. Radar and Sonar Systems}
		
		\parindent=0mm \vspace{0.1in}
		In radar and sonar signal processing, signals often undergo rapid frequency modulations, Doppler shifts, and transient behaviors due to environmental factors, moving targets, or changes in signal propagation. The Boostlet Transform can provide high-resolution time-frequency representations of these modulated signals, enabling improved detection, tracking, and classification of targets. Its ability to adapt the time-frequency resolution based on the signal's local characteristics makes it particularly valuable for analyzing chirp signals, frequency-hopping radar, and other systems where frequency changes over time.
		
		\parindent=0mm \vspace{0.1in}
		{\it 5.2. Communications Systems}
		
		\parindent=0mm \vspace{0.1in}
		In modern communication systems, especially in frequency-hopping and spread-spectrum communications, signals often change frequency in a non-linear and time-varying manner. The Boostlet Transform is well-suited to track these variations with improved precision compared to traditional methods like the Short-Time Fourier Transform (STFT). It can be used for analyzing signals in systems such as OFDM (Orthogonal Frequency Division Multiplexing), CDMA (Code Division Multiple Access), and adaptive modulation schemes, where the frequency content dynamically shifts in response to changing channel conditions. This enables better optimization of channel usage and improved detection and decoding of modulated signals.
		
		\parindent=0mm \vspace{0.1in}
		{\it 5.3. Speech and Audio Processing}
		
		\parindent=0mm \vspace{0.1in}
		Speech and audio signals are highly non-stationary, with rapidly changing frequency content during phoneme transitions or musical note changes. The Boostlet Transform can be applied to speech recognition, music analysis, and audio compression to achieve better time-frequency resolution during transient events, such as plosives or musical articulation. Its ability to adapt to the local frequency characteristics of a signal allows for more accurate representation of these non-stationary features, which is critical in improving the performance of speech-to-text systems, sound source separation, and other audio processing tasks.
		
		\parindent=0mm \vspace{0.1in}
		{\it 5.4.  Seismic and Geophysical Signal Processing}
		
		\parindent=0mm \vspace{0.1in}
		Seismic signals, such as those recorded in oil exploration, earthquake monitoring, and mining, often exhibit non-stationary behavior, with abrupt changes in frequency content corresponding to different geological layers or seismic events. The Boostlet Transform can be employed to analyze seismic data by adapting the time-frequency window to the specific characteristics of the signal, providing improved detection of transient seismic events and enhancing the interpretation of seismic waves. Its flexibility can help identify important features such as fault lines, underground cavities, or other geophysical phenomena that require precise time-frequency analysis.
		
		\parindent=0mm \vspace{0.1in}
		{\it 5.5.  Time-Frequency Imaging and Signal Detection}
		
		\parindent=0mm \vspace{0.1in}
		In imaging applications, such as medical ultrasound, optical coherence tomography, and non-destructive testing, time-varying signals are often used to probe materials or tissues. The Boostlet Transform can improve the resolution and accuracy of these imaging systems by providing better time-frequency analysis of the reflected signals, enhancing feature extraction and localization of key structures or anomalies. This leads to better-quality images and more precise diagnostics.
		
		\parindent=0mm \vspace{0.1in}
		{\it 5.6. Non-Stationary Signal Classification}
		
		\parindent=0mm \vspace{0.1in}
		In signal classification, particularly for signals that are non-stationary or exhibit complex temporal behavior, the Boostlet Transform can be used to better capture the evolving characteristics of the signal. For instance, in biometric signal analysis (e.g., gait analysis, heart rate variability), gesture recognition, or motion detection, the Boostlet Transform’s adaptability enables more precise feature extraction, leading to improved classification accuracy. This is crucial in applications like security systems, human-computer interaction, and health monitoring.

		\parindent=0mm\vspace{0.2in}
		{\bf{Declarations}}
		
		\parindent=0mm\vspace{0.1in}
		{\bf{Data Availability}} As no datasets were created or examined for this research, data sharing is not applicable.

		\parindent=0mm\vspace{0.1in}
		{\bf{Conflict of Interest}} The author declares that there are no competing interests.
		
		\parindent=0mm\vspace{0.2in}
		{\bf{{References}}}

\end{document}